\documentclass[aps,reprint,amsmath,amssymb,
 floatfix,superscriptaddress]{revtex4-2}

\usepackage{graphicx}
\usepackage{bm}
\usepackage{braket}
\usepackage{enumitem}
\usepackage{hyperref}
\usepackage{natbib}

\begin{document}
\title{One-dimensional repulsive Hubbard model with mass imbalance: Orders and filling anomaly}

\author{Yuchi He}
\affiliation{Department of Physics, Carnegie Mellon University, Pittsburgh, Pennsylvania 15213, USA}
\affiliation{Pittsburgh Quantum Institute, Pittsburgh, Pennsylvania 15260, USA}
\affiliation{Institute for Theory of Statistical Physics, RWTH Aachen University, and JARA Fundamentals of Future Information Technology, 52062 Aachen, Germany}
\author{David Pekker}
\affiliation{Pittsburgh Quantum Institute, Pittsburgh, Pennsylvania 15260, USA}
\affiliation{Department of Physics and Astronomy, University of Pittsburgh, Pittsburgh, Pennsylvania 15260, USA}
\author{Roger S. K. Mong}
\affiliation{Pittsburgh Quantum Institute, Pittsburgh, Pennsylvania 15260, USA}
\affiliation{Department of Physics and Astronomy, University of Pittsburgh, Pittsburgh, Pennsylvania 15260, USA}

\begin{abstract}
We investigate the phase diagram of the one-dimensional repulsive Hubbard model with mass imbalance. Using DMRG, we show that this model has a ``triplet" paired phase (dubbed $\pi \mathrm{SG}$) at generic fillings, consistent with previous theoretical analysis. 
We study the topological aspect of $\pi \mathrm{SG}$ phase,
determining long-range string orders and the filling anomaly which refers to the relation among the single particle gap, inversion symmetry, and filling imbalance for open chains. We also find, using DMRG, that at $1/3$ filling, commensurate effects lead to two additional phases: a crystal phase and a trion phase; we construct a description of these phases using Tomonaga-Luttinger liquid theory.

\end{abstract}

\maketitle

\section{Introduction}
 Mixtures of particles with different masses are indeed ubiquitous in nature. A minimal quantum model of such an interacting system is the mass-imbalanced Hubbard model.
By tuning the mass ratio from one to zero, this model bridges two limits: the extensively studied mass-balanced Hubbard model~\cite{essler2005one, Zheng1155} and the Falicov-Kimble model~\cite{Gruber1994, PhysRevLett.117.146601}.   The realization of the mass-imbalanced Hubbard model and its continuous counterpart~\cite{PhysRevLett.100.010401, PhysRevA.98.063624, ravensbergen2019strongly}, using cold atoms and other quantum simulators has motivated theoretical studies of its phase diagrams~\cite{fath1995asymmetric, cazalilla2005two, PhysRevLett.96.190402,mathey2007commensurate, PhysRevLett.103.105304,roux2011multimer,GubbelsStoofphysrep,PhysRevA.91.053611, Zdybel_2018}. 
Experimentally, the development of correlation measurements~\cite{PhysRevA.81.031610, Endres200, PhysRevLett.107.103001, mazurenko2017cold, Hilker484, Salomon2018, Chiu251} contributes to the exploration of correlated phases. Correlation measurements are particularly useful for detecting quasi-long-range orders of one-dimensional quantum phases of cold atom systems; measuring string orders also becomes possible~\cite{Endres200, Hilker484, Salomon2018}. This enables experimental observations~\cite{Hilker484, Salomon2018, Vijayan186} of the exactly solvable one-dimensional Hubbard physics.

There have been several theoretical studies of the one-dimensional mass-imbalanced Hubbard model. Although the model is no longer exactly solvable, it is expected to be described  within the framework of the Tomonaga-Luttinger (TL) theory. It has been shown that introducing mass imbalance leads to rich phase diagrams~\cite{cazalilla2005two,mathey2007commensurate,roux2011multimer}.
In particular, for repulsive interactions,  theoretical analyses predict that introducing mass imbalance opens a ``spin gap" once the Fermi vectors of the two components match,  even if the mass imbalance is infinitesimal~\cite{cazalilla2005two, mathey2007commensurate}. (The model does not have physical spin. The term ``spin" stems from the analogy to the two components of spin half.) However, the predicted spin-gapped liquid has not been observed numerically~\cite{roscilde2012pairing}.  On the other hand, crystallization was found at both 1/2~\cite{PhysRevB.97.085152} and 1/3 filling~\cite{roscilde2012pairing}.

In this paper, we revisit the phase diagram of the repulsive, mass-imbalanced Hubbard model.  We demonstrate the existence of a spin gap by studying its two-point correlations and string orders. We find novel aspects of the filling anomaly for systems with open boundaries, that is, the ground state has one more heavy particle than light particle, a feature protected by the spin gap.

We first show that by DMRG data analyses, the phase diagram can be confidently constructed, and the result is consistent with the TL theory.

For  ``incommensurate" fillings, i.e., the filling ratio is irrational, we find and characterize the spin-gapped phase, which we call $\pi \mathrm{SG}$ phase because the spin boson is locked at the value $\pi$ in the bosonized language. This confirms the prediction for the generic phase at equal-filling with a minor but interesting caveat we will discuss. We use various ways to characterize the $\pi \mathrm{SG}$ phase. Although a definite numerical resolution of a spin gap is demonstrated at relatively large mass imbalance and interaction, our data analysis techniques make it possible to indicate the existence of a spin gap at relatively small imbalance, and the result is not inconsistent with that a spin gap exists even at infinitesimal imbalance. 

For simple commensurate fillings, i.e., the filling ratio with a small denominator, we confirm that in addition to the $\pi \mathrm{SG}$ phase, it's possible to get crystals and liquids of bound states composed by particle(s) from one component and hole(s) from the other.  As an example, for one-third filling, we numerically show that the liquid phase of bound-states of two heavy holes with one light particle (named as trion) and $\pi \mathrm{SG}$ phase are separated by a crystal phase in the phase diagram.
We point out the relation among the three phases can be interpreted through TL theory: the locking of the spin boson gives $\pi \mathrm{SG}$, while the locking of another linear combination of boson fields gives trion; the locking of both fields leads to the formation of a crystal phase. Binding different numbers of particles and/or holes, like trion here,  are suggested to feature fractional conductance in DC transport~\cite{fractionalconductance1D,fractionalconductance1Dexp}.

Another goal of this paper is to figure out the novel aspects of $\pi \mathrm{SG}$ phase, due to the spin gap together with the mass imbalance. There has been work on the $\pi \mathrm{SG}$ phase~\cite{GiamarchiZ2fermion, Berg1DTSCz} on systems with additional spinful time-reversal symmetry. Due to the  mass imbalance, the spinful time-reversal symmetry is absent in our model; we highlight some of the characteristics which are similar and those that differ.

The $z$-component ``spin" density wave (``SDWz") is a characterization of the $\pi \mathrm{SG}$ phase because the phase difference of density-wave quasi-long-range order of each component is locked to be $\pi$. Without spinful time-reversal symmetry, there can be a difference between the two density-wave quasi-long-range order amplitudes,  leading to the coexistence with a total charge density wave (CDW). Connecting this observation with TL theory,  we illustrate that the form of bosonization representation of operators may depend on the symmetry of the Hamiltonian.

The $\pi \mathrm{SG}$ phase with additional spinful time-reversal symmetry has been considered to be topological, with characterizations of edge modes~\cite{Berg1DTSCz, PhysRevB.97.115107} and string orders~\cite{verresen2019gapless}.  For the mass-imbalanced Hubbard model, spinful time-reversal symmetry is absent while inversion symmetry survives.  In this case, we find the string order structures~\footnote{Long-range string order is known to be absent for bosonic gapped inversion symmetry protected Haldane phase~\cite{PhysRevB.81.064439,PhysRevB.85.075125}; by contrast, the fermion parity symmetry makes long-range string order possible for gapped and even gapless fermionic systems~\cite{perez2008string, PhysRevLett.109.236404, PhysRevLett.113.267204, emtll, PhysRevB.99.115113}.} remain robust while more possibilities are opened for edge physics, summarized as ``filling anomaly". Via bosonization, we figure out that the long-range string order structures are constrained by the inversion symmetry.  For open chains, the ``fourfold degeneracy" in the spin sector protected by spinful time-reversal symmetry is lifted in our case. The remaining feature is that we have a spin-gapped ground state  with  one more heavy particle than the light particles. This is one possibility of 
the filling anomaly, a terminology means that spin gap, inversion symmetry, and  filling balance cannot be realized simultaneously. (The terminology is borrowed from a similar phenomenon illustrated in free fermion systems in Refs.~\onlinecite{PhysRevB.99.245151,khalaf2019boundary}.)  The phenomenon is explained in terms of bosonization.

\section{Hamiltonian and methods}\label{HM}
\begin{figure}[tb]
(a) \raisebox{-0.9\height}{\includegraphics[width=8cm]{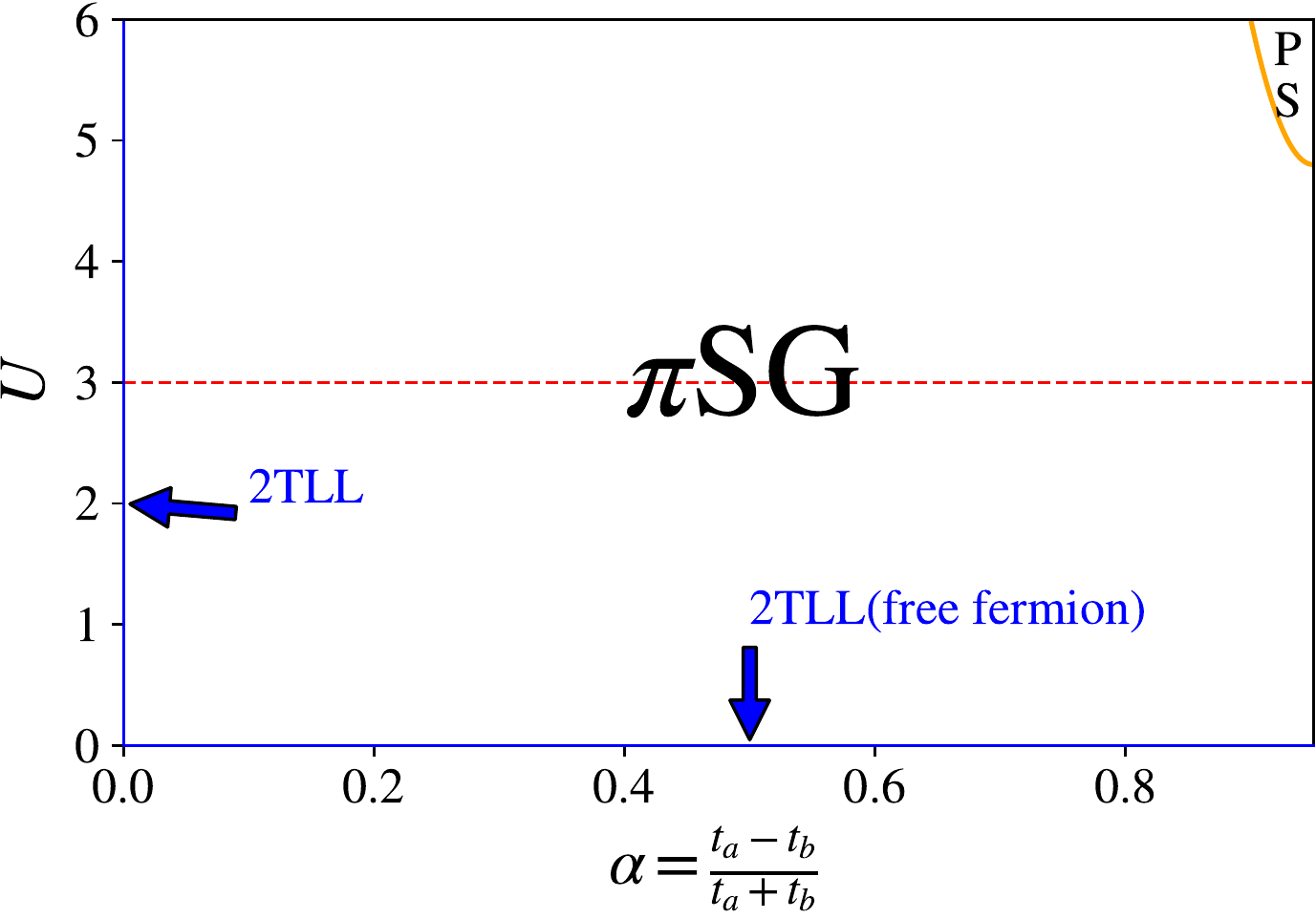}}
\\
(b) \raisebox{-0.9\height}{\includegraphics[width=8cm]{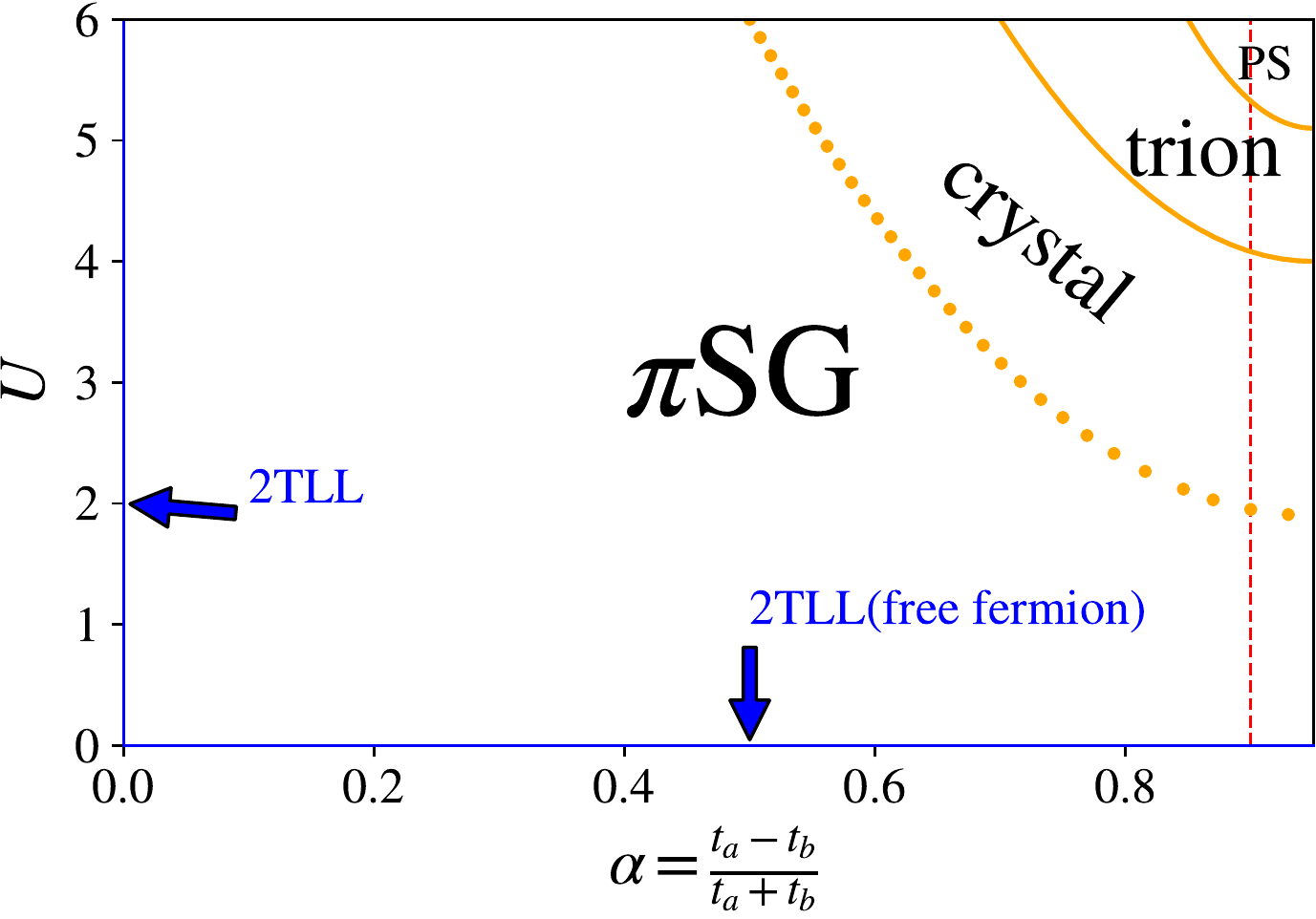}}
\caption{Phase diagram of the model Eq.~\eqref{eq:Hubbard} at (a) ``incommensurate" fillings: (b) one-third filling.
''PS" denotes phase separated.
The red dashed lines denote the cuts of data we present in Figs.~\ref{fig:xi} and~\ref{fig:onethirdxi}. The phase boundaries (orange lines) are estimated based on similar analysis illustrated in Fig.~\ref{fig:xi} and Fig.~\ref{fig:onethirdxi}. The phase boundary between $\pi \mathrm{SG}$ and crystal (dotted line) is approximate because it is difficult to extract numerically. 2TLL phase is likely to only exist either in the mass balance ($\alpha=0$) or non-interacting ($U=0$) limit.}\label{fig:phase diagram}
\end{figure}

Consider the Hubbard Hamiltonian:
\begin{align}
H &= \sum_{\substack{x \\ \sigma\in\{a,b\}}} \left[\!\begin{array}{r} -t_\sigma \left(c^\dagger_{\sigma}(x) c_{\sigma}(x+1)+\text{h.c.}\right)
\;\\{} + U n_{a}(x) n_{b}(x) \end{array}\!\right], \label{eq:Hubbard}
\end{align}
where $c_{\sigma}(x)$ annihilates a $\sigma$ fermion at site $x$, $n_{\sigma}=c^{\dagger}_{\sigma}c_{\sigma}$. Mass imbalance means $t_a\neq t_b$. Through Jordan-Wigner transform, the model Eq.~\eqref{eq:Hubbard} is equivalent to its hardcore bosonic version~\cite{roscilde2012pairing} as well as fermion-hardcore boson mixture version~\cite{PhysRevLett.96.190402}.
We focus our studies on the ground state phases of Eq.~\eqref{eq:Hubbard} with $t_at_b>0$, $U>0$  and equal fillings $\left\langle n_a \right\rangle = \left\langle n_b \right\rangle$.
The phase diagram is affected by whether the filling is some simple fraction (i.e., the denominator of the irreducible fraction is small), where the commensurate effect can alter the phase even for moderate interaction. We start from the simpler case of incommensurate filling or negligible commensurate effect. Numerically, we must pick a commensurate(rational) filling; we perform DMRG calculation on 5/11 filling and find the results can represent  ``incommensurate" cases.  We then work with simpler fraction filling (1/2, 1/3) to discuss the commensurate effects. Estimated phase diagrams of  5/11 and 1/3 filling are summarized in Fig.~\ref{fig:phase diagram}.

We use both bosonization and DMRG to study the quantum phases.   Following the standard bosonization approach,  possible phases are constructed as descendant phases of the 2TLL (two-component Tomanaga-Luttinger liquid) phase, which is adiabatically connected to the non-interacting limit.  Whether a phase indeed exists in a given parameter region can be answered by DMRG.  We use  infinite DMRG (iDMRG)~\cite{mcculloch2008infinite,iDMRGpaper} to calculate the phase diagrams Fig.~\ref{fig:phase diagram}, we fix $t_a+t_b=2$ and $t_a>t_b>0$. We define the ``imbalance parameter'' $\alpha=\frac{t_a-t_b}{t_a+t_b}$ with $0\leqslant\alpha<1$.
Our phase diagrams are parameterized by $\alpha$ and $U$. The finite DMRG has also been implemented, only for studying edge effects and extracting spin gaps (Figs.~\ref{fig:phizfinite},~\ref{fig:spingap} and~\ref{fig:crystalspingap}).

The elementary gapless charge excitations of all observed phases are different from each other. Notice that the particle number of each component conserves separately, and thus the charge carried by an excitation (local operator) is denoted as $(q_a,q_b)$.  The sector of any integer multiple of the elementary gapless charge is  also gapless.
The two-component TLL (2TLL) phase has two types of elementary gapless charge excitations with charge $(1,0)$ and $(0,1)$, which are the single particle excitations of the two components. The $\pi \mathrm{SG}$ has an odd parity (``triplet") pairing quasi-long-range order;  the elementary gapless charge excitations carry a charge $(1,1)$. The trion phase is a liquid of bound states of two heavy particles and one light hole, thus the elementary gapless excitation carries charge $(-1,2)$. The crystal phase, on the other hand, does not have gapless excitations.
In our context, the central charge ($c$) of a phase equals the number of elementary gapless charge excitation types.

The gaplessness in a charge sector indicates the corresponding correlation length is divergent. The correlation length $\xi_{q_a,q_b}$  is defined as the maximal length scale of correlators taking form $\braket{A^\dag(0)B(r)}$ where $A,B$ are charge-$(q_a,q_b)$ operators. It is convenient to read out the estimation of correlation length from iDMRG. The iDMRG generates infinite matrix product states (iMPS) as the approximation of the ground states. The correlation lengths of an iMPS can never be divergent. But one can increase the number of variational parameters, which is characterized by bond dimension ($\chi$), to infer if $\xi_{q_a,q_b}(\chi)$ is divergent in the infinite-$\chi$ limit. Such a limit is believed to capture the exact ground states. The data of finite-$\chi$ correlation lengths is illustrated for the two cuts of the phase diagrams in Fig.~\ref{fig:xi} and~\ref{fig:onethirdxi}.  A theoretical understanding of how the $\xi_{q_a,q_b}(\chi)$ is supposed to diverge with $\chi$~\cite{pollmann2009theory} is very helpful for inferring divergence or convergence.  In short, we pick the bond dimensions in a geometric series, i.e., $\chi=625, 1000, 1600, 2560$. For a divergent physical correlation length $\xi_{a,b}$, the series $\ln\xi_{a,b}(\chi)$ is expected to be equally-spaced if $\chi$ is large enough. The value of the spacing is related to the central charge of phases~\cite{pollmann2009theory}; the two reference spacings for central charge 1 and 2 are relevant to our analysis.

Besides $\xi_{q_a,q_b}(\chi)$, the data of orders and string orders at finite bond dimensions can also provide information. In the large $\chi$ limit, long-range (string) orders are characterized by a non-zero value. Under some condition, quasi-long-range orders can be characterized by power law decaying of long-range orders with $\xi_{q_a,q_b}(\chi)$, where $\xi_{q_a,q_b}(\chi)$ diverges in the large $\chi$ limit. Scaling dimensions can be extracted from the powers. The idea is used to extract the Luttinger parameter and provide evidence for small spin gaps.

Detailed discussion of technical aspects is provided in Appendix~\ref{DMRGnotes}. Technical aspects can be skipped for understanding the main story of this paper, i.e., the orders, quasi-long-range orders, strings orders, and filling anomaly of those strongly correlated phases.

\section{``incommensurate" filling: \texorpdfstring{$\pi \mathrm{SG}$}{piSG} phase}~
We start our discussion at incommensurate filling  to avoid the additional complexity of commensurate effects. Previous theoretical studies~\cite{cazalilla2005two, mathey2007commensurate} predict a spin-gapped phase, which we call $\pi \mathrm{SG}$ phase, as the only stable phase. In this section, we first revisit the basics of $\pi \mathrm{SG}$ phase from TL theory and then provide numerical analyses of several physical quantities to identify and study the $\pi \mathrm{SG}$ phase.  In discussing the numerical analyses, we will interlude with some theoretical analyses to show how $\pi \mathrm{SG}$ with mass imbalance differs from its spinful time-reversal symmetric counterpart.

We use $\phi_a$ and $\phi_b$ to bosonize $c_a$ and $c_b$ respectively:
\begin{align}
c_{\sigma}(x) &= \frac{\kappa_{\sigma,+}}{\sqrt{2\pi}} e^{i[\theta_{\sigma}+(\phi_{\sigma}+k_{\text{F},\sigma}x)]} + \frac{\kappa_{\sigma,-}}{\sqrt{2\pi}} e^{i[\theta_{\sigma}-(\phi_{\sigma}+k_{\text{F},\sigma}x)]}
\notag\\&\quad + \dots,      
\end{align}
where $\kappa_{\sigma, \pm}$ is Klein factors and $\theta_{\sigma}$ is the dual field of $\phi_{\sigma}$.  
It is helpful to introduce a set of rotated basis for the two boson fields:
\begin{align}\label{chargespin}
\text{charge boson:}\quad& \phi_c=\frac{1}{\sqrt{2}}(\phi_a+\phi_b),\\
\text{spin boson:}\quad& \phi_s=\frac{1}{\sqrt{2}}(\phi_b-\phi_a).
\end{align}
If the low energy physics is described by a quadratic Lagrangian  of $\phi_{\sigma}$ and $\theta_{\sigma}$, the ground state is a 2TLL. The descendant phases of 2TLL are constructed by locking vertex terms. We consider terms invariant under inversion $x \rightarrow -x$, $\phi_{\sigma} \rightarrow -\phi_{\sigma}$ together with homogeneity. In this case, the most relevant locking (vertex) term, representing an inter-component backward scattering, is $g_s\cos(2\sqrt{2}\phi_s)$ with $g_s>0$ expected for repulsive interactions in our model.
The $\pi \mathrm{SG}$ phase is constructed by locking the spin boson $2\sqrt{2}\phi_s$ at (odd multiples of) $\pi$. The label $\pi$ before $SG$ (spin-gapped) distinguishes the phase with the other spin-gapped phase with locking value $\braket{2\sqrt{2}\phi_s} = 2\pi \times \text{integer}$. 
 $\pi \mathrm{SG}$ ($2\pi \mathrm{SG}$) locking minimizes $g_s\cos(2\sqrt{2}\phi_s)$ for $g_s>0 \ (<0)$. We note that expectation values of vertex operators are useful for determining the nature of the quasi-long-range order. For example, \ for the $\pi \mathrm{SG}$ phase, $\phi_s=\pi/(2\sqrt{2})$. Hence for $m \in \mathbb{Z}$, $\left\langle\cos((2m+1)\sqrt{2}\phi_s)\right\rangle =\left\langle\sin(2m\sqrt{2}\phi_s)\right\rangle=0$ while  $\left\langle\sin((2m+1)\sqrt{2}\phi_s)\right\rangle \neq 0$, $\left\langle\cos(2m\sqrt{2}\phi_s)\right\rangle \neq 0$. Notice that the two locking values give distinct phases is provided by the fact that $\cos(2\sqrt{2}\phi_s+\delta)$ term with continuous $\delta$ is disallowed. As discussed, this can be protected by inversion symmetry  along with homogeneity.

\subsection{Bulk spin gaps and  string orders}\label{pilocking}
In this subsection, we confirm the existence of the $\pi \mathrm{SG}$ phase by looking at its defining property: charge excitation is gapless; spin excitation is gapped due to a $\pi$-locking. We use correlation length analysis to infer whether the charge and spin excitation is gapped, we then propose that string orders can confirm the $\pi$-locking and show our numerical results.

\begin{figure}
\includegraphics[width=0.4\textwidth]{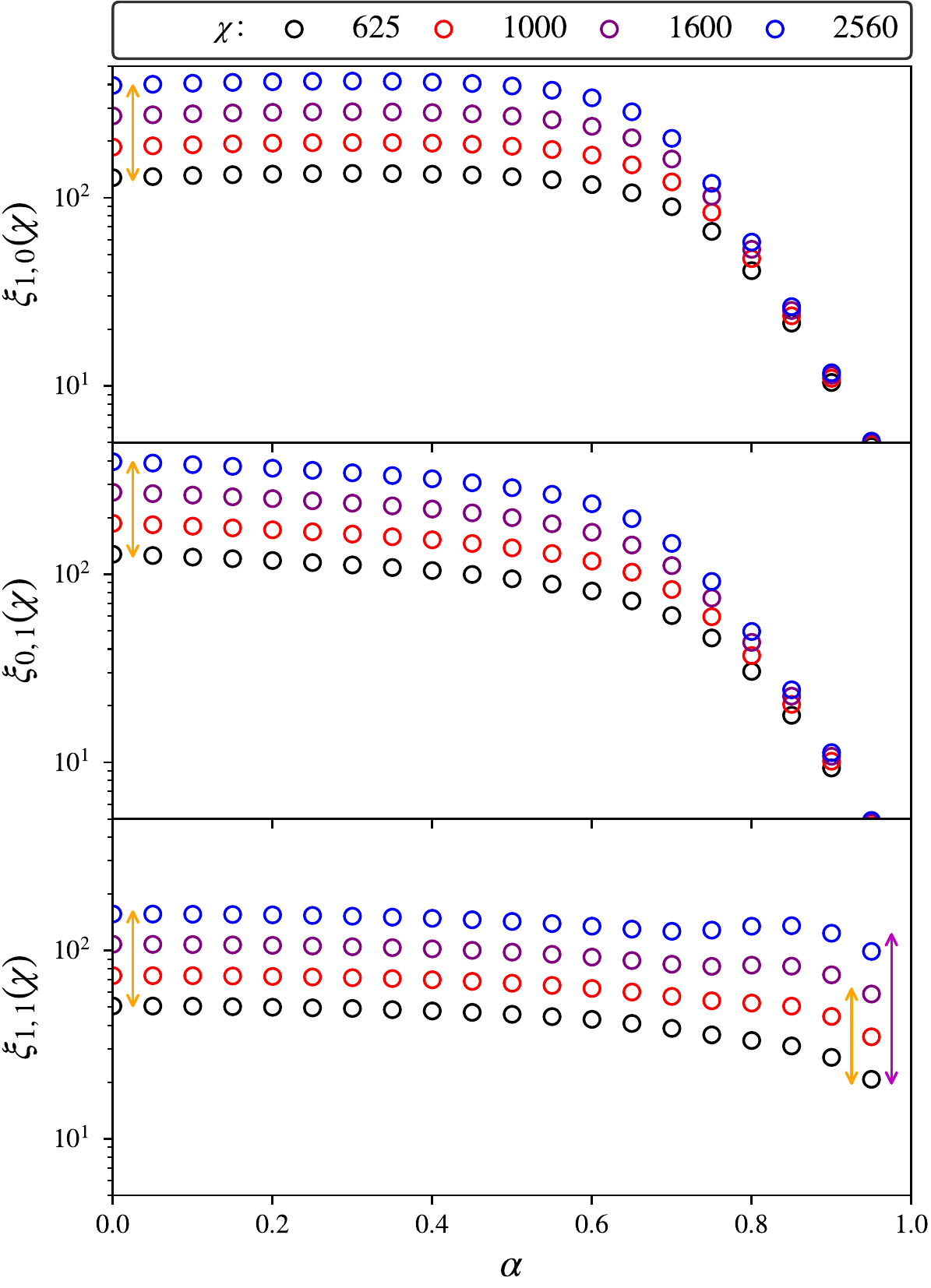}
\caption{Finite-$\chi$ correlation length as a function of mass imbalance $\alpha=\frac{t_a-t_b}{t_a+t_b}$ along the cut $U=3$ at 5/11 filling.
The correlation lengths in the (1,0)-, (0,1)-, and (1,1)-charge sectors are computed for various bond dimensions $\chi$.
A spin-gapped phase including $\pi \mathrm{SG}$ is characterized by finite physical $\xi_{1,0}$, $\xi_{0,1}$ but divergent $\xi_{1,1}$ (in the limit $\chi\to\infty$).
The data clearly shows that for $\alpha>0.75$, $\xi_{1,0}(\chi)$ and $\xi_{0,1}(\chi)$ tend to be convergent with increasing $\chi$ and hence a spin gap.
The solid orange and magenta lines represent the expected difference between $\ln(\xi(\chi=625))$ (black circle) and $\ln(\xi(\chi=2560))$ (blue circle) in a gapless sector of $c=2$ and $c=1$ systems respectively. We see that the numerical result near $\alpha=0$ is consistent with a 2TLL state ($c=2$).
For $\alpha \gtrsim 0.75$, the result (for $\xi_{1,1}$) clearly shows a $c=1$ phase. In Figs~\ref{fig:orderparameter} and ~\ref{fig:orderparameterscaling}, we show that another data analysis of the same set of wave-function data extends the evidence of spin gaps for smaller $\alpha$.
}
\label{fig:xi}
\end{figure}

We first show that the spin boson is indeed gapped out while the charge boson is gapless. This is to say that the unit charge of gapless excitations is (1,1). To show this, we use iDMRG to demonstrate that correlation lengths $\xi_{1,0}$ and $\xi_{0,1}$ are finite while the correlation length $\xi_{1,1}$ diverges.
In Fig.~\ref{fig:xi}, we plot the three correlation lengths through a cut of data with $U=3$. In this plot, the convergence (divergence) can be inferred through finite-$\chi$ scaling, which we have explained in Sec.~\ref{HM}.
Recall, for the approximations of a ground state (a vertical cut), the spacing of $\xi$ between points with neighboring values of $\chi$ indicates whether the charge sector is gapped and/or the central charge of the theory. 
In particular, for a gapless sector $Q$ of $c=1$ (or 2), we expected $\ln(\xi_Q(\chi=1000))-\ln(\xi_Q(\chi=625)) \approx 0.632$ (or 0.409), while for a gapped sector $\Delta\ln\xi$ converges to zero as $\chi$ increases.
For large mass imbalance, $\alpha>0.75$, we observe that $\xi_{1,0}(\chi)$ and $\xi_{0,1}(\chi)$ tend to be convergent with increasing $\chi$, while $\xi_{1,1}(\chi)$ tends to be divergent. The convergence tendency gradually becomes less clear for $\alpha<0.75$.
We plot the reference $\xi$ increment as vertical lines for the gapless sectors of $c=1$ (magenta) and 2 (orange) from $\chi=625$ (bottom circle) to $\chi=2560$ (top circle), $\ln(\xi(\chi=2560))-\ln(\xi(\chi=625))\approx 3 \times 0.632 \ (3 \times 0.409)$. Recall that at $\alpha=0$, we have 2TLL phase with $c=2$ and the orange line of increment is consistent with the data. On the other hand, the magenta line is close to the increment of $\xi_{1,1}$ at large $\alpha$, indicating a $c=1$ phase.
As explained in Appendix~\ref{DMRGnotes}, the data analysis of $\xi$ dependence on $\chi$ at a single value of $\alpha$ cannot distinguish if the spin gap is strictly zero or small but finite. This is indeed the situation for small $\alpha$'s.  We will return to this issue in Sec.~\ref{spingapprevalence} and use other data analysis techniques to show evidence of spin gaps for smaller $\alpha$.

We now show that the spin boson $2\sqrt{2}\phi_s$ is locked at odd multiples of $\pi$. We define the following string lattice operators~\cite{stringcorrelator, verresen2019gapless} which can serve as order parameters to distinguish two types of locking. Those operators are:
\begin{subequations}\label{Jz}
\allowdisplaybreaks
  \begin{align}
   \Phi_{n}(x) &\equiv \left[\prod_{j<x}(-1)^{n(j)}\right] (1-2n(x)) \nonumber \\
&=\frac{1}{2}\left[\prod_{j<x} Q(j) \right] (Q_a(x)+Q_b(x)); \label{phin}\\
\Phi_{z}(x) &\equiv \left[ \prod_{j<x}(-1)^{n(j)} \right] \sigma_{z}(x) \nonumber  \\
&=\frac{1}{2}\left[ \prod_{j<x} Q(j)\right](Q_a(x)-Q_b(x)),
  \label{phiz}\end{align}
\end{subequations}
where $\sigma_z=n_b-n_a$, $n=n_{a}+n_{b}$, and $Q_\sigma=(1-2n_{\sigma})$. We let $Q=Q_aQ_b$ be the fermion parity operator. For infinite system, the correlation functions rather than the expectation values of $\Phi_n$ and $\Phi_z$ are well defined; the values $\left\langle \Phi_{n}(0) \Phi_{n}(\infty) \right\rangle $, $\left\langle \Phi_{z}(0) \Phi_{z}(\infty) \right\rangle$ are concerned.

Based on the considerations we will soon discuss, we claim that  bosonization representation of $\Phi_n$ and $\Phi_z$ are:
\begin{subequations}\label{bosonization}
\begin{align}
\Phi_{n}(x) &\sim \cos(\sqrt{2}\phi_{s})+...,\label{bosonizationa}\\
\Phi_{z}(x) &\sim \sin(\sqrt{2}\phi_{s})+...\label{bosonizationb}
\end{align}
\end{subequations}
With Eq.~\eqref{bosonization}, $2 \sqrt{2}\phi_s$ being locked at $\pi$ dictates that  $\Phi_{z}(x)$ (Eq.~\eqref{bosonizationb}) is ordered while $\Phi_{n}(x)$ (Eq.~\eqref{bosonizationa}) is disordered.

We provide a physical picture of why  $\Phi_{z}(x)$ is ordered in the $\pi \mathrm{SG}$ phase. This picture is the ``squeezed space" interpretation of $\Phi_z$~\cite{stringcorrelator, Hilker484}. The squeezed space is constructed  by excluding the empty and double occupied sites and relabelling the remaining single-occupied sites as sites in a (shortened) chain~\cite{stringcorrelator}.
In the squeezed space, one can thus define an effective spin-1/2 model.
 $\Phi_z$ can be considered as the N\'eel order parameter in the squeezed space.
In our model, N\'eel order $\Phi_z$ forms in the squeezed space. It is made possible by on-site repulsion together with the absence of SU(2) symmetry.  We now draw an analogy between Eq.~\eqref{bosonizationb} and the bosonization of a spin-1/2 chain. Note that if we define $\tilde{\phi}_s=\phi_s/\sqrt{2}$, we obtain $\Phi_{z}(x) \sim \sin(2\tilde{\phi}_s)$. This is the bosonization representation of N\'eel order of a spin chain with the convention that $\sigma_z \sim \frac{1}{\pi}\partial_x \tilde{\phi}_s$. The ``$\tilde{\phi}_s$" in the squeezed space is related to $\tilde{\phi}_s$  by re-scaling the length~\cite{stringcorrelator}.  

Appendices~\ref{spatialparitystring}   --~\ref{bosonizationapp} show Eq.~\eqref{bosonizationa} and provide an alternate argument for Eq.~\eqref{bosonizationb} by considering inversion symmetry and commutation relations.

\begin{figure}[htbp]
\includegraphics[width=8cm]{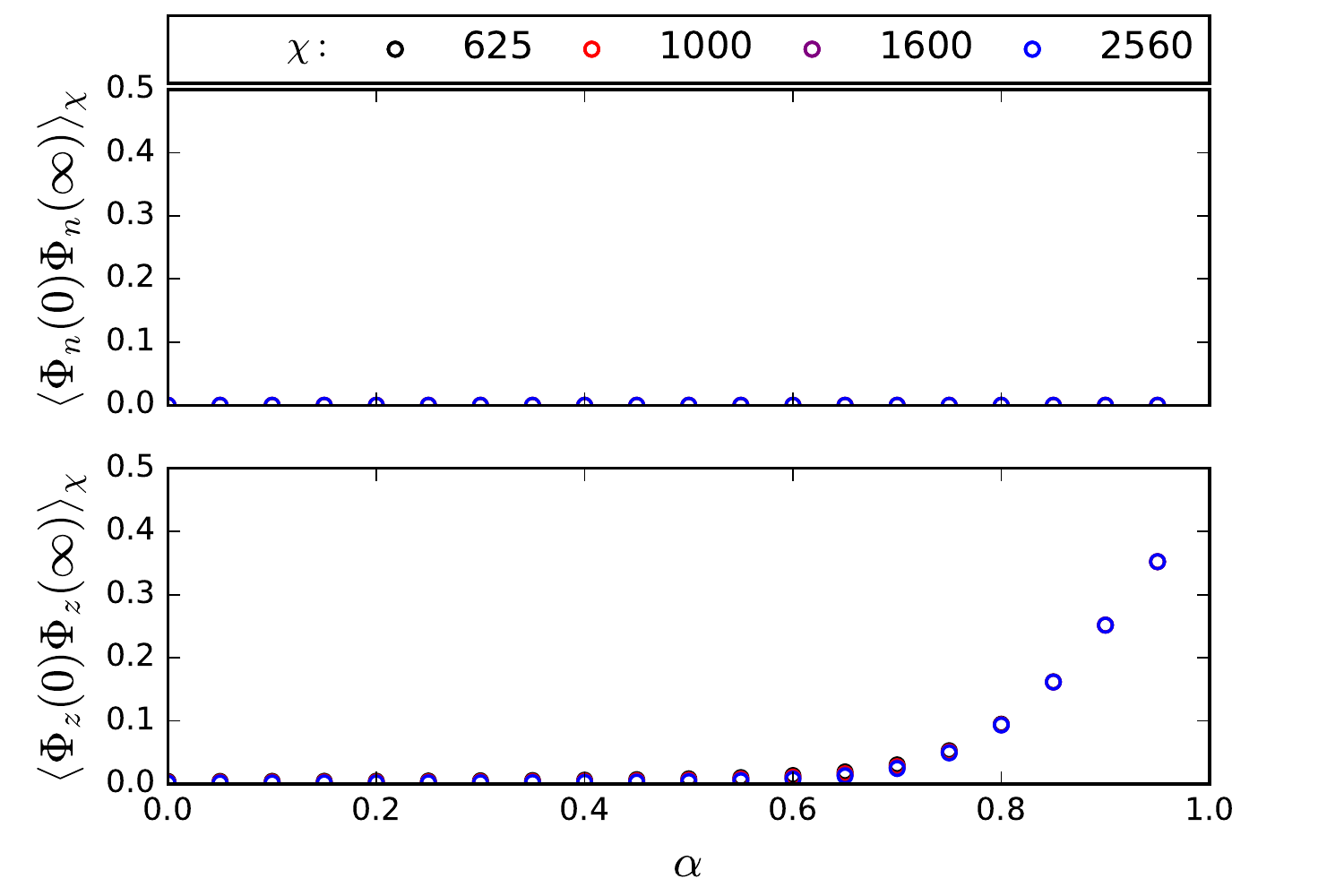}
\caption{Order parameter characterization of $\pi \mathrm{SG}$ states. The parameters of the data points are the same as  Fig.~\ref{fig:xi} ($U=3$ at filling 5/11). The data is obtained by unit cell averaging which partially eliminates the finite bond dimension effects. Top panel shows that $\left\langle \Phi_{n}(0) \Phi_{n}(\infty) \right\rangle =0$ (at the accuracy of order $<10^{-10}$). Bottom panel demonstrates the finite expectation values of $\left\langle \Phi_{z}(0) \Phi_{z}(\infty) \right\rangle$. The data point $\alpha=0.9$ can be compared to our finite DMRG result in Fig.~\ref{fig:phizfinite} and the value is consistent.  }\label{fig:orderparameter}
\end{figure}

Evaluating string orders for our data with $U=3$, we find that $\left\langle \Phi_{n}(0) \Phi_{n}(\infty) \right\rangle =0$ while $\left\langle \Phi_{z}(0) \Phi_{z}(\infty) \right\rangle \neq 0$, as predicted for $\pi \mathrm{SG}$ phase,  see Fig.~\ref{fig:orderparameter}.  We also verify that for $2\pi \mathrm{SG}$ phase obtained by setting $U<0$ in our model, $\Phi_{z}(x)$ is disordered and $\Phi_{n}(x)$ is ordered. For both phases,  we find $\left\langle \Phi_{n}(0) \Phi_{z}(\infty) \right\rangle =0$, consistent with Eq.~\eqref{bosonization}.  Consider an alternative basis for the string order parameters Eq.~\eqref{Jz}: $\Phi_a=\frac{1}{2}\left[\prod_{j<x} Q(j)\right]Q_a(x)$ and $\Phi_b=\frac{1}{2}\left[\prod_{j<x} Q(j)\right]Q_b(x)$. We have $\left\langle \Phi_a(0)\Phi_a(\infty)\right\rangle= \left\langle \Phi_b(0)\Phi_b(\infty)\right\rangle=-\left\langle \Phi_a(0)\Phi_b(\infty)\right\rangle$ for $\pi \mathrm{SG}$ phase; for $2\pi \mathrm{SG}$ phase, we find $\left\langle \Phi_a(0)\Phi_a(\infty)\right\rangle= \left\langle \Phi_b(0)\Phi_b(\infty)\right\rangle=+\left\langle \Phi_a(0)\Phi_b(\infty)\right\rangle \neq 0$. Such
result, consistent with Eq.~\eqref{bosonization}, appears surprising, because the components $a$ and $b$ are not symmetric in our case. Nevertheless, the $a$, $b$ exchange symmetry of the string correlations, only emerges at the long-distance limit.

\subsection{String order and filling anomaly in open chains}\label{open chains}
\begin{figure}[tbh]
\includegraphics[width=\columnwidth]{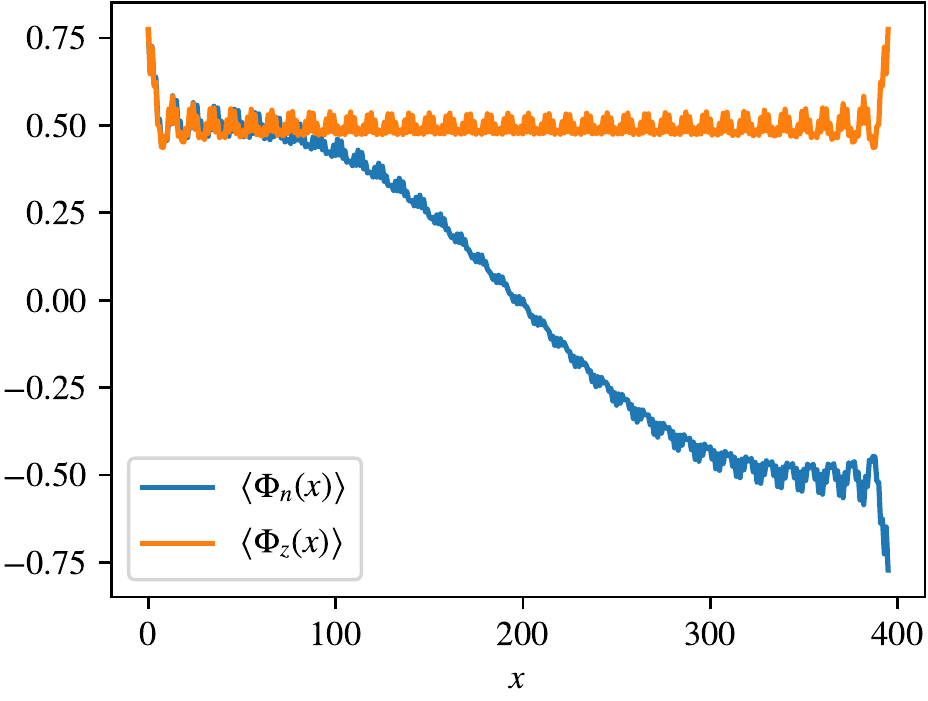}
\caption{The string order $\left\langle \Phi_n(x) \right\rangle$, $\left\langle \Phi_z(x) \right\rangle$ of a finite chain with length $L=396$, 179 light ($a$ type)  particles and 180 heavy ($b$ type)  particles. The other parameters for this plot are $U=3, \alpha=0.9$.  In the bulk,  $\left\langle \Phi_z(x) \right\rangle$ is locked around a finite value, which is quantitatively consistent with the iDMRG result (Fig.~\ref{fig:orderparameter}) of string correlation. 
}
\label{fig:phizfinite}
\end{figure}

\begin{figure}
(a) \raisebox{-0.9\height}{\includegraphics[width=8cm]{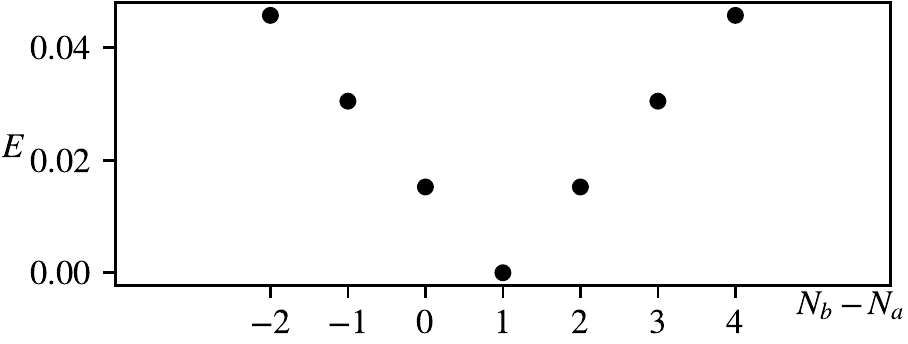}}
\\
(b) \raisebox{-0.9\height}{\includegraphics[width=8cm]{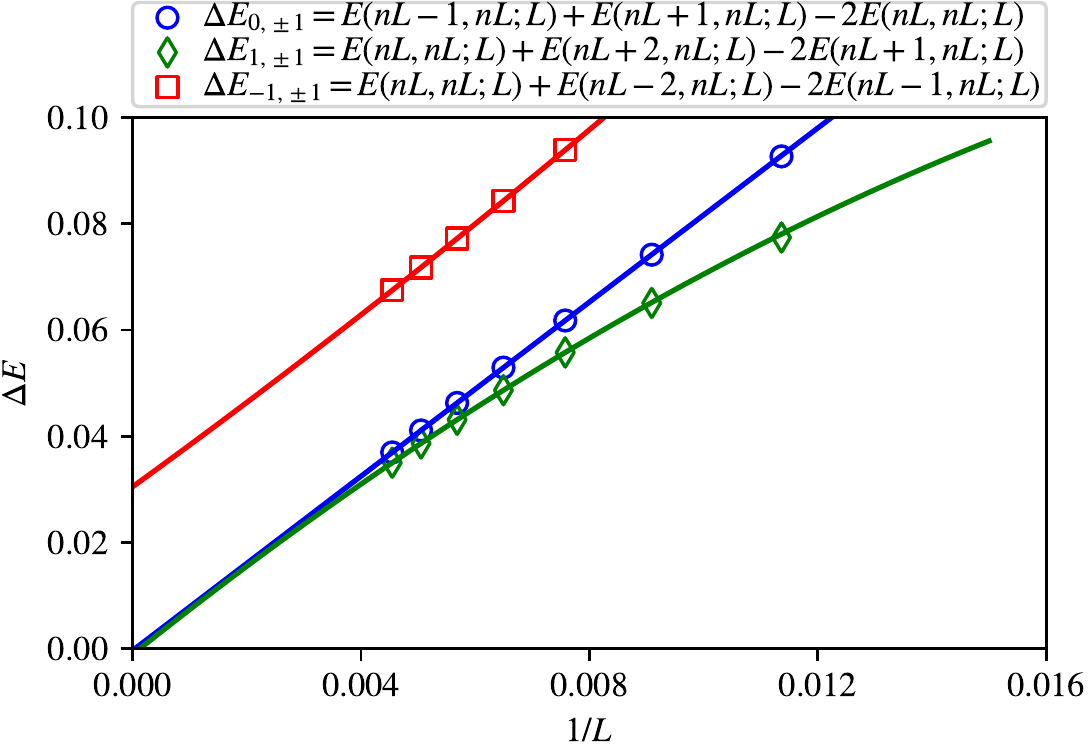}}
\caption{Excitation energies for open chains as a function of particle numbers.
(a) The energy of lowest-energy state within each sector with fixed particle numbers; the size of the open chain is extrapolated to infinity; the chemical potential is chosen to make the data left-right symmetric.
(b)  The spin gap is estimated by fitting the intercept ($\Delta E(\frac{1}{L}=0)$) of $\Delta E(\frac{1}{L})$ curve.  The $E(N_a, N_b; L)$ used to define $\Delta E(\frac{1}{L})$, is the ground state energy of the open chain with length $L$, $N_a$ light particles and $N_b$ heavy particles.  The curve is obtained from a polynomial fit up to  $1/L^2$. The quantity $\Delta E(\frac{1}{L}=0)$ is twice the spin gap.  We see a finite intercept ($\approx 0.03$) of the upper curve comparing to the vanishing intercepts of the other two ($\approx -0.001, -0.0003 $).  The two small negative but finite intercepts are likely to mostly come from the fitting ansatz/ finite size errors. If using polynomial fit up to  $1/L^3$, the two  intercepts shrink to $\approx -8 \times 10^{-5}$; the raw DMRG energy data accuracy is at the order $10^{-7}$.  The  parameters of these plots: $U=3$, $\alpha=0.9$ and macroscopic density $n=5/11$.
}
\label{fig:spingap}
\end{figure}

Next, we discuss $\pi \mathrm{SG}$ phase with open boundaries. Usually when there is a spin gap and absence of disorder, one expects the ground state to have equal fillings---i.e., $N_a=N_b$---for a finite range of chemical potentials.
Here, we show that for open chains, the $\pi \mathrm{SG}$ phase can lead to  one-particle filling imbalance, which is the key signature of the filling anomaly.
We first use effective theory to explain the filling anomaly phenomenon  from the configuration of spin boson.
We provide our numerical data for string orders (Fig.~\ref{fig:phizfinite}) consistent with the proposed field configuration.
We then give direct evidence for filling anomaly (Fig.~\ref{fig:spingap}) by measuring the spin gap.
We discuss the filling anomaly in the $\pi \mathrm{SG}$ phase when there is an additional spinful time-reversal symmetry.

\begin{figure}[tbh]
\includegraphics[width=\columnwidth]{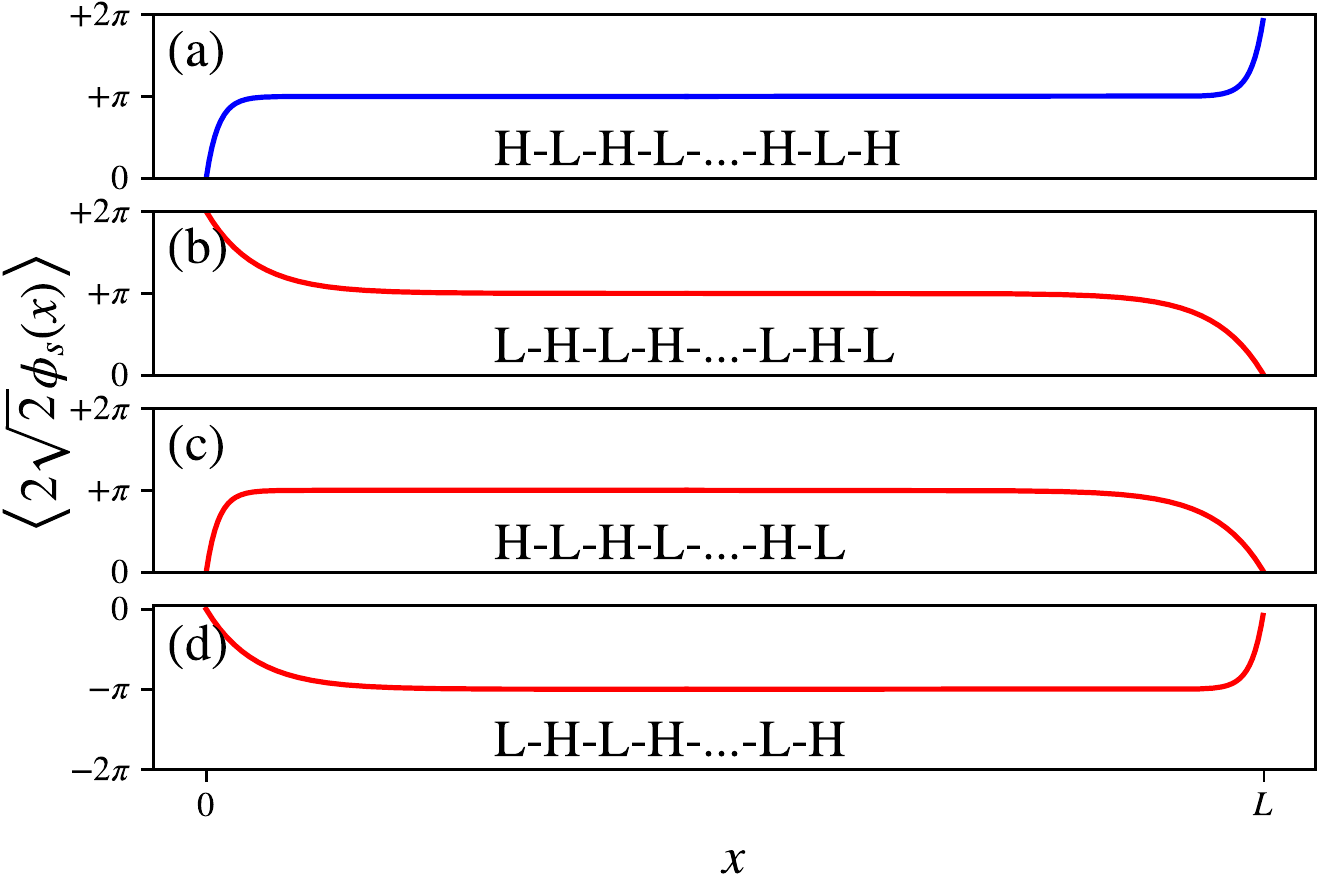}
\caption{
(a) Schematic plot of the spin boson field configuration of the observed spin-gapped state with open boundary condition. The height difference between right and left indicates that there is one more heavy (type $b$) particle than the light (type $a$) particle. This one more particle is localized at the two edges as  $\sigma_z \sim \frac{\sqrt{2}}{\pi}\partial_x \phi_s$. The locking of $\phi_s$ in the bulk leads to the locking of the microscopic string operator $\Phi_z$ in the bulk, e.g., \ Fig.~\ref{fig:phizfinite}.  (b)--(d) Some alternative configurations with a spin gap. It turns out that they do not  describe observed $\pi \mathrm{SG}$ ground states. The (c)--(d) describe two assumed states at precise equal filling with inversion symmetry spontaneously broken. For (a)--(d), the strings of H (heavy particle) and L (light particle) denote the N\'eel order in the squeezed space.}
\label{fig:gsspinbosonconfig}
\end{figure}

First, we discuss the possible spin boson field configurations for a state in $\pi \mathrm{SG}$ phase with open boundary conditions.  
The name of $\pi \mathrm{SG}$ phase comes from that the field $2\sqrt{2}\phi_s$ is locked at $\pi$ in the bulk. For infinite chains, the locking value is defined modulo  $2\pi$. For open chains, however, those values are not equivalent as boundary condition needs to be considered. The boundary condition is that in the left and right vacuum, $2\sqrt{2}\phi_s$ is locked at integer multiples of $2\pi$. Without losing generality, we fix $2\sqrt{2}\phi_s$ of the left vacuum at 0~\cite{Berg1DTSCz}.
The locking value difference of right and left as multiples of $2\pi$ counts the number difference of the two types of particles, see  Appendix~\ref{bosonizationsymmeries} for details.
The fact that the bulk locking value is not an integer multiple of $2\pi$ indicates that there must be a change of field expectation value near the edges. As the spin-$z$ density is related to the spin boson $\sigma_z \sim \frac{\sqrt{2}}{\pi}\partial_x \phi_s+...$, there must be non-zero particle number imbalance near each edge.
Four spin field configurations could be relevant to stable ground states, plotted in Fig.~\ref{fig:gsspinbosonconfig}. According to the previous discussion, configurations (a) and (b) have one more heavy and light particle, respectively. Configurations (c) and (d) are particle number balanced but there is ``spin" polarization near the edges which spontaneously breaks the inversion symmetry. [Notice that the states described by Figs.~\ref{fig:gsspinbosonconfig}(a) and (b) are inversion symmetric, as the spin boson itself is not an observable, but its derivative is a component of the ``spin" imbalance operator.] Intuitively, the configuration (a) with one heavy particle localized at edges is more likely to be energetically stable than those with one or half a light particle  localized near the edge(s). This is because the kinetic energy contribution of the localized heavy particle can be smaller.

The assumption that only field  configuration (a) is stable results in two predictions: the long-range string order $\Phi_z(x)$ only exists in the sector with one more heavy particle; the spin gap only locks the filling to that with one more heavy particles for a finite range of chemical potentials. We show that our numerical data is consistent with the predictions.  We use the string operators $\Phi_z(x)$ and $\Phi_n(x)$ to probe the expectation value of the spin boson $\phi_s$ in the bulk (Fig.~\ref{fig:phizfinite}); we also  compute the spin gap and show that the state with one additional heavy particle is separated with other spin sectors by a finite gap and nearby sectors are in the continuum (Fig.~\ref{fig:spingap}).

First, we measure the string operators on the ground state with one more heavy particle. Fig.~\ref{fig:phizfinite} plots $\langle \Phi_z(x) \rangle$ and $\langle \Phi_n(x) \rangle$ for given parameters. 
According to Eq.~\eqref{bosonization}, the leading contribution to these string operators can indicate the locking of the spin boson.
We observe that there is a plateau for $\langle \Phi_z(x) \rangle$ in the bulk. It is clear that the result is consistent with the spin boson locked at $\pi$ and the $\sin(\sqrt{2}\phi_s(x))$ term in the bosonization representations. Here, $\sin(\sqrt{2}\phi_s(x))$ is expected to lock at a non-zero value in the bulk as $2\sqrt{2}\phi_s(x)$ is locked at $\pi$.
We observe that $\langle \Phi_n(x) \rangle$ does not  have a finite locking value, as  $\cos(\sqrt{2}\phi_s(x))$ vanishes in the bulk.  This is consistent with the requirement of odd inversion parity of the curve.  The curve $\Phi_n(x)$ is ``soft" in the bulk, which can be encoded by those terms with charge mode in the bosonization expansions, in addition to Eq.~\eqref{bosonization}, see Appendix.~\ref{bosonizationapp}.  From the definition Eq.~\eqref{Jz}, we observe that $\Phi_z(x)$ becomes $\sigma_z(x)=\frac{1}{2}(n_b(x)-n_a(x))$ at the left and right edges; $\Phi_n(x)$ becomes $\frac{1}{2}(1-n_b(x)-n_a(x))$ and $\frac{1}{2}(n_a(x)+n_b(x)-1)$ at the left and right edges respectively. This explains the feature near the edges. We also observe Friedel oscillations. Those features can be encoded by other terms of the bosonization expansions.

Next, we measure the ground state energy for each spin sector and extract the spin gap.  Fig.~\ref{fig:spingap}(a) plots the ground state energy within each spin sector, relative to the absolute ground state energy. The data has been extrapolated to the thermodynamic limit.  We see the data points can be connected by two straight lines, which intersect at one kink; the kink is located at $N_b=N_a+1$ (when there is an additional heavy particle).  Since adjusting the relative chemical potentials tilts the plot, the existence of the kink ensures that for a finite range of chemical potentials, the ground state has the filling $N_b-N_a=1$.
We define $\Delta E$ as the sum of the energy cost of adding and subtracting a light ($a$ type) particle respectively. “$\Delta E$ is the difference of the right and the left slope.  $\Delta E$ is invariant  under a change of chemical potentials.  The spin gap is defined as  half of $\Delta E$. The extrapolation of $\Delta E$ to thermodynamic limit is demonstrated in Fig.~\ref{fig:spingap}(b).
The data indicates that $\Delta E$ is non-zero only if the base state is chosen in the spin sector that there is one more heavy particle ($N_b=N_a+1$).   This is consistent with that only one sector has a bulk spin gap. The value of $\Delta E$ is approximately $0.03$ for the particular parameters while the corresponding  bulk single particle correlation length is $\sim 10$ (see Sec.~\ref{pilocking}).
Here, the spin gap is evaluated by subtracting and adding one light particle. A consistent result is obtained by subtracting and adding one heavy particle.    We remark that there is no low energy edge mode.  The existence of such mode is not consistent with the observed energy landscape, i.e. crossing of the two straight lines [Fig.~\ref{fig:spingap}(a)]. Thus, $\Delta E$ indeed reflects the bulk spin gap.

The phenomenon that there is one more heavy particle in the spin-gapped state is one possibility of filling anomaly which is a more general statement of topological edge effects than edge zero modes. In this case, the non-local inversion symmetry protection can be involved,  comparing to symmetry with locality like spinful time-reversal symmetry. Similar to Refs.~\onlinecite{PhysRevB.99.245151, khalaf2019boundary}, we may state that for an open chain without impurity, the ground state(s) with a bulk $\pi$ spin gap, there is either inversion symmetry breaking or particle number imbalance.  Keselman and Berg~\cite{Berg1DTSCz} have studied open  $\pi \mathrm{SG}$ chain with extra spinful time-reversal symmetry with a discussion of spin boson field configurations.    In their case, the time-reversal symmetry action on spin boson configuration: $\phi_s(x) \rightarrow -\phi_s(x)$ can transform Fig.~\ref{fig:gsspinbosonconfig}(a) into Fig.~\ref{fig:gsspinbosonconfig}(b).  This means the slope amplitude of the configuration should be the same, unlike what we have drawn for the lack of spinful time-reversal symmetry.  By doing the transform at one of the edges, the other two ground states similar to (c) and (d) with equal particle numbers can be obtained. This procedure does not change the energy in the thermodynamic limit, due to locality and spinful time-reversal symmetry.  While being equal-filling, these two states break both spinful time-reversal and inversion symmetry.  Hence, their four degenerate ground states all fit in the statement of  filling anomaly.
In our case, spinful time-reversal symmetry is absent; so the simultaneous existence of spin gaps in the three sectors is not guaranteed. Our calculation indicates that the only spin-gapped state is in the sector with one more heavy particle; the gapless equal-filling states do not spontaneously break inversion symmetry.  Our result that the spin-gapped state is inversion symmetric but has one more heavy particle also fits in the statement of filling anomaly.

\subsection{Quasi-long-range orders}\label{qlo}
In the following paragraphs, we show how to characterize the $\pi \mathrm{SG}$ phase using spin, charge, and pair quasi-long-range orders in infinite systems. Specifically, we focus on the decay exponents and oscillatory wavevectors of the algebraic decaying components of the various correlators. These exponents, in turn, all depend on the single Luttinger parameter $K$ in the effective bulk theory Eq.~\eqref{chargesectoreft}:
\begin{align}\label{chargesectoreft}
	\mathcal{H}_{c} &= \frac{v_{c}}{2\pi}\bigg[K(\partial_{x}\theta_{c})^2 + \frac{1}{K}(\partial_{x}\phi_{c})^2\bigg].
\end{align}

For the correlation of operators in the neutral (0,0) sector, we now show that mass imbalance mixes the quasi-long-range orders of ``spin" and charge. The bosonization representation of the density of  species $\sigma$ is:
\begin{align}\label{bosonizationdensity}
n_{\sigma}=\frac{1}{\pi}\partial_x \phi_{\sigma}+\lambda_{\sigma}\sin(2\phi_{\sigma}+2k_{\text{F}}x)+..., 
\end{align}
where $\lambda_{\sigma}$ is non-universal~\cite{gogolin2004bosonization,LUKYANOV1998533}. As mass imbalance introduces an asymmetry between $a$ and $b$ components, we expect $\lambda_a \neq \lambda_b$ generically for interacting systems.  Then using Eqs.~\eqref{chargespin} and~\eqref{bosonizationdensity}, we write the bosonization representations of $n$ and $\sigma_z$:
\begin{subequations}
\begin{align}
   n&=\frac{\sqrt{2}}{\pi}\partial_x \phi_{c}+(\lambda_{a}+\lambda_{b})\sin(\sqrt{2}\phi_c+2k_{\text{F}}x)\cos(\sqrt{2}\phi_{s})\nonumber \\ 
&\quad+(\lambda_{b}-\lambda_{a})\cos(\sqrt{2}\phi_c+2k_{\text{F}}x)\sin(\sqrt{2}\phi_{s}) +\dots,\label{bosonizationn}\\
\sigma_z&=\frac{\sqrt{2}}{\pi}\partial_x \phi_{s}+(\lambda_{a}+\lambda_{b})\cos(\sqrt{2}\phi_c+2k_{\text{F}}x)\sin(\sqrt{2}\phi_{s}) \nonumber\\
&\quad+(\lambda_{b}-\lambda_{a})\sin(\sqrt{2}\phi_c+2k_{\text{F}}x)\cos(\sqrt{2}\phi_{s}) + \dots.\label{bosonizationsigmaz}
\end{align}
\end{subequations}

A consequence of $\lambda_{a} \neq \lambda_{b}$ is that the density-wave quasi-long-range orders of the ``spin" and charge should qualitatively be the same, i.e., only differ in their coefficients.  In this sense, even with a spin gap, there is no strict spin-charge separation.
The leading quasi-long-range orders for $n$ and $\sigma_z$ are: 
\begin{align}
\left\langle n(0) n(x)\right\rangle \sim \left\langle \sigma_z(0) \sigma_z(x)\right\rangle \sim \frac{\cos(2k_{\text{F}}x)}{|x|^{K}}.
\end{align}
This is different from the mass-balanced case, in which the $2k_{\text{F}}$ quasi-long-range order of either ``spin" or charge vanishes. That case is a consequence of setting $\lambda_{a}=\lambda_{b}$ in Eqs.~\eqref{bosonizationn}~\eqref{bosonizationsigmaz} and plugging in the $\phi_s=\pi/(2\sqrt{2})$ ($\pi \mathrm{SG}$ phase) or $\phi_s=0$ ($2\pi \mathrm{SG}$ phase).

\begin{figure}[tb]
\includegraphics[width=\columnwidth]{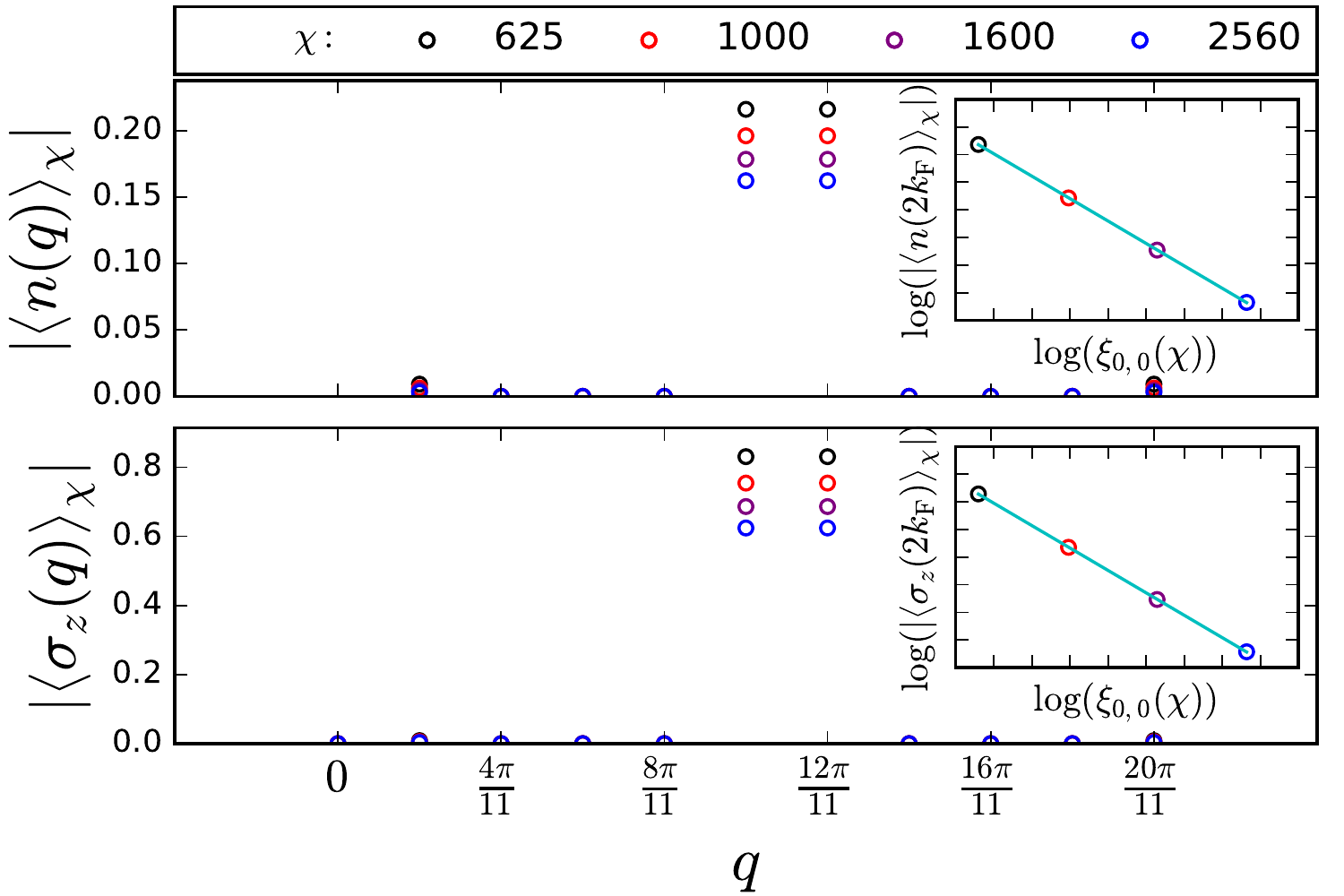}
\caption{Scaling of finite-$\chi$ charge-density-wave order (top panel) and spin-density-wave order (bottom panel) of a $\pi \mathrm{SG}$ state with bond dimension $\chi$  and finite-$\chi$ correlation length $\xi_{0,0}(\chi)$. The two orders show peaks at $q=\frac{10\pi}{11}$ and $q=\frac{12\pi}{11}$ which correspond to $q = 2k_{\text{F}}$ and $q = 2\pi-2k_{\text{F}}$. The two insets show how the $2k_{\text{F}}$ peaks decay with $\xi_{0,0}(\chi)$ as we tune the bond dimension $\chi$. 
We estimate that the Luttinger parameter $K\approx0.406$ by fitting the slope in the inset and using the proposition that $\text{slope}=-K/2$. The  point $|\langle n(q=0) \rangle_{\chi}|$, which is fixed by the total charge density, is not plotted.  Plot parameters: $U=3$, $\alpha=0.8$, and filling $5/11$.}
\label{fig:densityorder}
\end{figure}

To extract the period and the exponent of the quasi-long-range order of the neutral sector, we study how finite-$\chi$ order decays with the DMRG correlation length as we increase $\chi$ (see Sec.~\ref{HM}). The decay exponent of the peak at $q=\pm 2k_{\text{F}}$ gives the scaling dimension of the leading order: $|\left\langle n(2k_{\text{F}}) \right\rangle_\chi| \sim |\left\langle \sigma_z(2k_{\text{F}}) \right\rangle_\chi| \sim \xi_{0,0}^{-K/2}$. In Fig.~\ref{fig:densityorder}, we plot $|\left\langle n(q) \right\rangle_\chi|$ and $|\left\langle \sigma_z(q) \right\rangle_\chi|$ of a $\pi \mathrm{SG}$ state and show the fitting of $K$.

\begin{figure}[tb]
(a) \raisebox{-0.9\height}{\includegraphics[width=7cm]{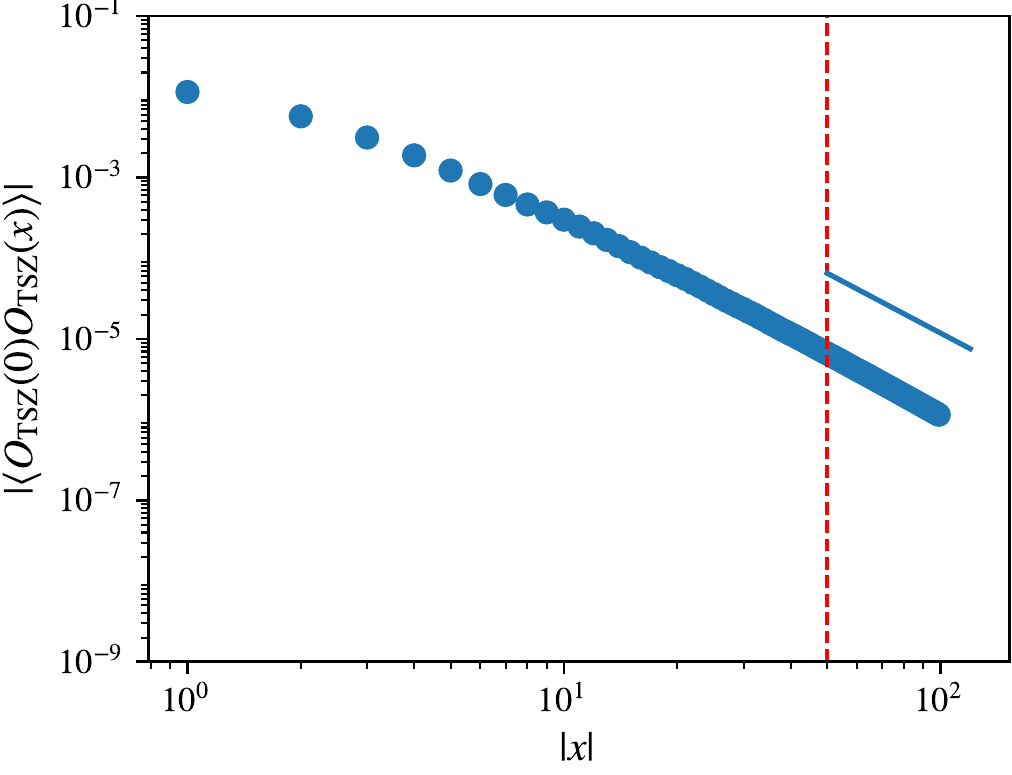}}
\\
(b) \raisebox{-0.9\height}{\includegraphics[width=7cm]{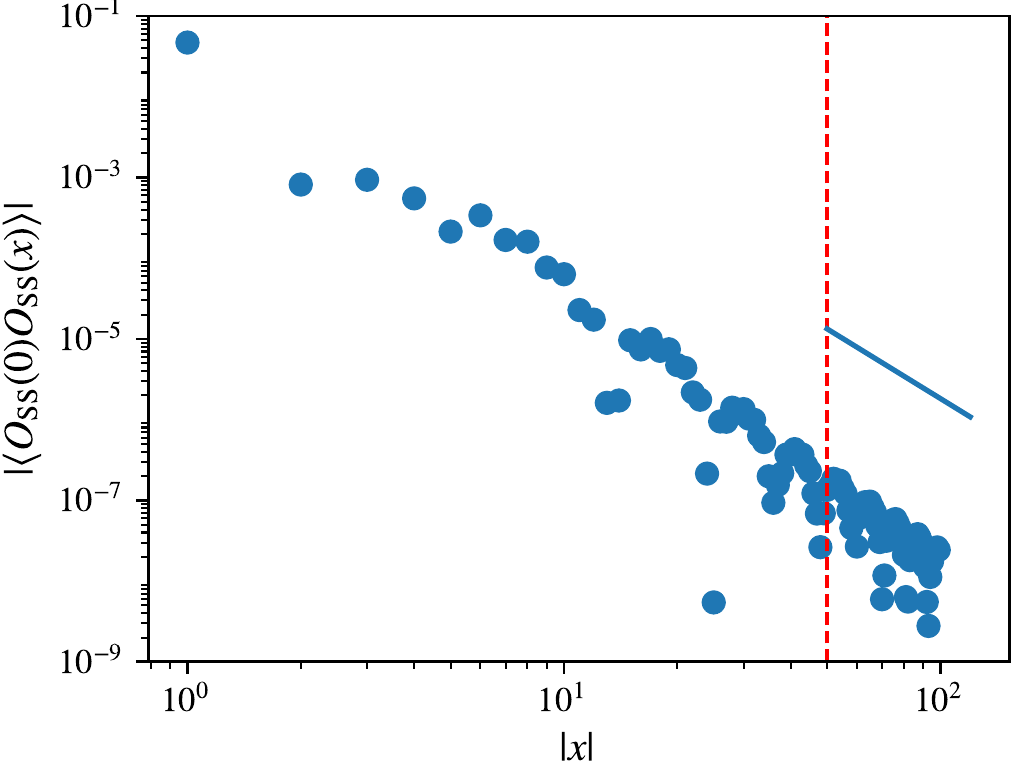}}
\caption{Pairing quasi-long-range orders of a state in $\pi \mathrm{SG}$ phase. The parameters of the state: $U=3$, $\alpha=0.8$, and filling $5/11$. The asymptotic behavior of the TSCz order (a) is predicted as $\sim 1/r^{1/K}$; while for the ``s-wave" order (b) is $\cos(2k_{\text{F}}x)/r^{1/K+K}$. $K$ can be independently extracted by SDWz/CDW order (Fig.~\ref{fig:densityorder}). Using the extracted value 0.406, we plot $1/r^{1/K}$ and $1/r^{K+1/K}$ (the solid lines) in the first and second figure respectively, approximately matching the leading decaying slope of the numerical data.  Note that the asymptotic behavior should be found at  distance beyond the correlation length of the spin boson (estimated in the (1,0) and (0,1) sectors and denoted by the red dashed lines).}
\label{paircorrelators}
\end{figure}

The leading quasi-long-range order in the pair sector (1,1) is the $z$-component ``triplet" pairing (TSCz) order~\cite{cazalilla2005two,mathey2007commensurate}.  The system neither has an SU(2) nor spinful time-reversal symmetry but has an inversion symmetry; the TSCz order here inherits the odd spatial parity property of TSCz in the context of SU(2) symmetry.  A microscopic representation of the order is
$O_{\mathrm{TSZ}}(x)=c_{a}(x)c_{b}(x+1)+c_{b}(x)c_{a}(x+1)$; the leading term of its two-point correlator is $\sim \frac{1}{|x|^{1/K}}$. Consider an ``s-wave" pair operator $O_{\mathrm{SS}}(x)=c_{a}(x)c_{b}(x)$, which has even parity;  TL theory predicts that the corresponding leading term of its two-point correlator is $\sim \frac{\cos(2k_{\text{F}}x)}{|x|^{1/K+K}}$, namely a pairing-density-wave quasi-long-range order. Notice that this pairing-density-wave quasi-long-range order in the ``s-channel" decays faster than the TSCz quasi-long-range order. In Fig.~\ref{paircorrelators}, we show correlators of $O_{\mathrm{TSZ}}$ and  $O_{\mathrm{SS}}$ respectively for a state in the $\pi \mathrm{SG}$ phase. We also plot the reference slopes of the two pair correlators, obtained by theoretical prediction and $K$ extracted from the CDW/SDWz quasi-long-range order. The good agreement between the reference slopes and the observed slopes of the two pair correlators indicates that our description is consistent.

\subsection{Discussion of the  prevalence of spin gap over the phase diagram}\label{spingapprevalence}
Detecting smaller spin gaps requires larger system sizes and/or bond dimensions. A region with a finite but small spin gap in the phase diagram, may get missed in numerical detection as in Ref.~\cite{roscilde2012pairing}. Here, we introduce some data analysis techniques to mitigate the challenge of detecting small spin gaps. In this subsection, we discuss the prevalence of the spin gap based on our data for $\alpha > 0$ with our data analysis techniques.

The TL theory based on perturbative RG or large $U$ expansion has predicted~\cite{cazalilla2005two, mathey2007commensurate} that the spin gap is prevalent and the gap scales as
\begin{align}~\label{gapscaling}
E_{\text{sg}} \sim e^{-B /(\alpha U\sin(k_{\text{F}}))},
\end{align}
where $B$ is roughly a positive constant.  The factor $\sin(k_{\text{F}})$ in the exponent indicates that the finite spin gap is a pure many-body effect; its value vanishes at the few-body limit $k_{\text{F}} \to 0, \pi$.
The exponent is $\propto 1/\gamma$, with $\gamma$ representing the detuning parameter $\alpha$ or $U$. This is different from the scaling $\ln(E_{\text{sg}}) \propto 1/\sqrt{\gamma}$ near the BKT transition of the XXZ chain or the attractive Hubbard model.
Therefore, due to the scaling $\propto 1/\gamma$, the rate of increase of the gap $E_{\text{sg}}$ (away from $\gamma=0$) in our model is even slower than that of the  transition of XXZ chain or the attractive Hubbard model. In this regard, we expect a large region of $\alpha$ or $U$ near zero with a tiny spin gap.

As we have discussed at the beginning of Sec.~\ref{pilocking}, the simple finite-$\chi$ correlation lengths scaling (Fig.~\ref{fig:xi}) cannot determine whether there is an indeed finite spin gap for the region $0<\alpha<0.75$. The region  $0<\alpha<0.65$ is also unclear from the plot on the string order (Fig.~\ref{fig:orderparameter}).  In the following paragraphs, we show that further data analysis provides evidence for the prevalence of spin gap and Eq.~\eqref{gapscaling}.

\begin{figure}[htbp]
\includegraphics[width=8cm]{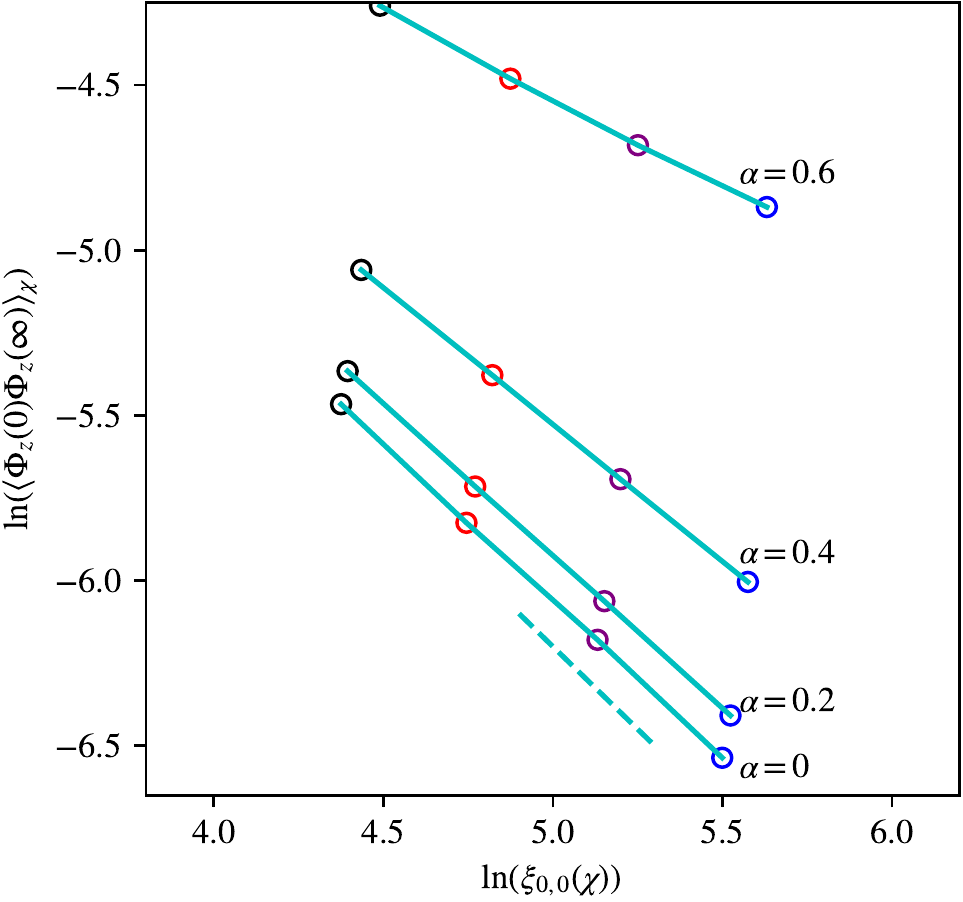}
\caption{Dependence $\left\langle \Phi_{z}(0) \Phi_{z}(\infty) \right\rangle_{\chi}$ for small $\alpha$ on DMRG correlation lengths $\xi_{0,0}(\chi)$. The dashed line is a reference of the slope $=1$, see the discussion in Sec.~\ref{spingapprevalence}.  }\label{fig:orderparameterscaling}
\end{figure} 

First, we use a scaling dimension analysis to argue that a spin gap persists for $\alpha>0$. We note that, for a 2TLL to be stable, we require the scaling dimension of $\Phi_z$, $\Delta_{\Phi_z} \geq 1/2$. This is because $\Delta_{\Phi_z}$ is a quarter of the dimension of the spin lock term ($\Delta_{\cos(2\sqrt{2}\phi_s)}$), which should be $\geq 2$ in the 2TLL phase. On the other hand, for in $\pi \mathrm{SG}$  phase, $\Phi_z$ takes expectation value,  $\Delta_{\Phi_z}=0$.
We numerically estimate $\Delta_{\Phi_z}$ by the finite-$\chi$ scaling (see Sec.~\ref{HM}) of $\left\langle \Phi_{z}(0) \Phi_{z}(\infty)\right\rangle_{\chi}$.
 The dimension $\Delta_{\Phi_z}$ is estimated as one half of the slope of the log-log plot of $\left\langle \Phi_{z}(0) \Phi_{z}(\infty)\right\rangle_{\chi}$ and $\xi_{0,0}(\chi)$. 

Fig.~\ref{fig:orderparameterscaling} shows our plot for $\alpha=0$, 0.2, 0.4, and $0.6$. In the large $\chi$ limit, the slope should be $\geq 1$ for 2TLL, and exactly $0$ for the $\pi \mathrm{SG}$ phase. At finite $\chi$, the estimated slope is found to be between 0 and 1, with larger mass imbalance $\alpha$ corresponds to a smaller slope. At the mass-balanced point ($\alpha=0$), the state has $\mathrm{SU}(2)$ symmetry which dictates that $\Delta_{\Phi_z}=1/2$; hence the asymptotic slope should be 1. Indeed, our data shows a good agreement; the estimated slope is 0.97 and tends to be closer to 1 for larger $\chi$.
For $\alpha=0.2, 0.4, 0.6$, the slopes at finite $\chi$ are clearly smaller than 1. The slopes for $\alpha=0.4, 0.6$ show a clear tendency to decrease for larger $\chi$, while such tendency is unclear for $\alpha=0.2$ (at $\chi \approx 2560$).
We remark that if there was a 2TLL phase at nonzero $\alpha$, we should have observed a slope greater than 1. Hence, the scaling dimension estimation provides evidence of the prevalence of the spin gap.

In the appendix~\ref{BKT scaling collapse}, we present the second analysis,  a BKT gap scaling collapse based on Eq.~\eqref{gapscaling}.

\section{commensurate filling}

For commensurate fillings, umklapp processes induce locking terms in the effective theory, which can be interpreted as binding between particle(s) and hole(s).  As a result, new phases may appear in the phase diagram, such as crystal phases or liquids of particle-hole bound states.
Hence, the TL analysis of the phase diagram should consider those terms in addition to the spin-locking term analyzed in the previous section.  In this section, we do indeed find new phases; we show that the phase diagrams can be understood by the competition and collaboration of umklapp term(s) with the spin-locking term.

The general form of interacting (vertex) terms is $\cos\left(2m\phi_a+2n\phi_b+(2m+2n)k_{\text{F}}x\right)$, where $m, n\in \mathbb{Z}$. Umklapp terms correspond to $m+n \neq 0$. In order for an umklapp term to lock, $2(m+n)k_{\text{F}}=2\pi N$,  where $N \in \mathbb{Z}$.  Thus, the choice $m$ and $n$ is restricted by the filling ($\text{filling}=k_{\text{F}}/\pi$). We note that except for half filling, no umklapp term can lock when the interaction is infinitesimal.  This is clear from scaling dimension of the Umpklapp term, which in the non-interacting limit is $m^2+n^2 \geq 2$, where the equality holds only for half filling (with $m=n=1$);  the critical scaling dimension is 2. 

For 1/2 filling,  the most relevant umklapp locking term is $g_c\cos(2\sqrt{2}\phi_c)$ with $g_c<0$. This term is believed to always lock in any stable phase of the repulsive model, consistent with the scaling dimension analysis.  A large charge gap can develop even for moderate $U \sim t_a, t_b$, and once a charge gap develops the spin sector can be analyzed separately.  For the case of a large charge gap, it has been shown theoretically that the spin gap can also open~\cite{fath1995asymmetric, cazalilla2005two}.   As a result, true long-range density order develops to form a crystal.  The long-range SDWz order in the crystal phase has been found numerically in Refs.~\onlinecite{fath1995asymmetric, PhysRevLett.96.190402}. Here, we observe that the long-range SDWz order coexists with the CDW order in this phase, see Appendix~\ref{crystalhalffilling}. While the existence of CDW order is not obvious from the infinite $U$ coupling analysis presented in Refs.~\onlinecite{fath1995asymmetric,cazalilla2005two}, it becomes obvious following our analysis Eq.~\eqref{bosonizationdensity},~\eqref{bosonizationn}, and~\eqref{bosonizationsigmaz}. Comparing to 5/11 filling, we see that in the half filling case, an easily detectable (and larger) spin gap extends over a broader range in parameter space. Whether the crystal phase is the only stable phase throughout the phase diagram at half filling remains an open question. To test this, one can perform scaling collapse. However, the possible direct BKT transition from the 2TLL to the crystal phase involves two lengths scales: one for the charge sector and one for the spin sector, which is numerically challenging. This type of two-parameter collapse of numerical data was performed in a completely different model~\cite{Shao213}, and hence could be possible in the future for our model.

\begin{figure*}[t]
\includegraphics[width=0.7\textwidth]{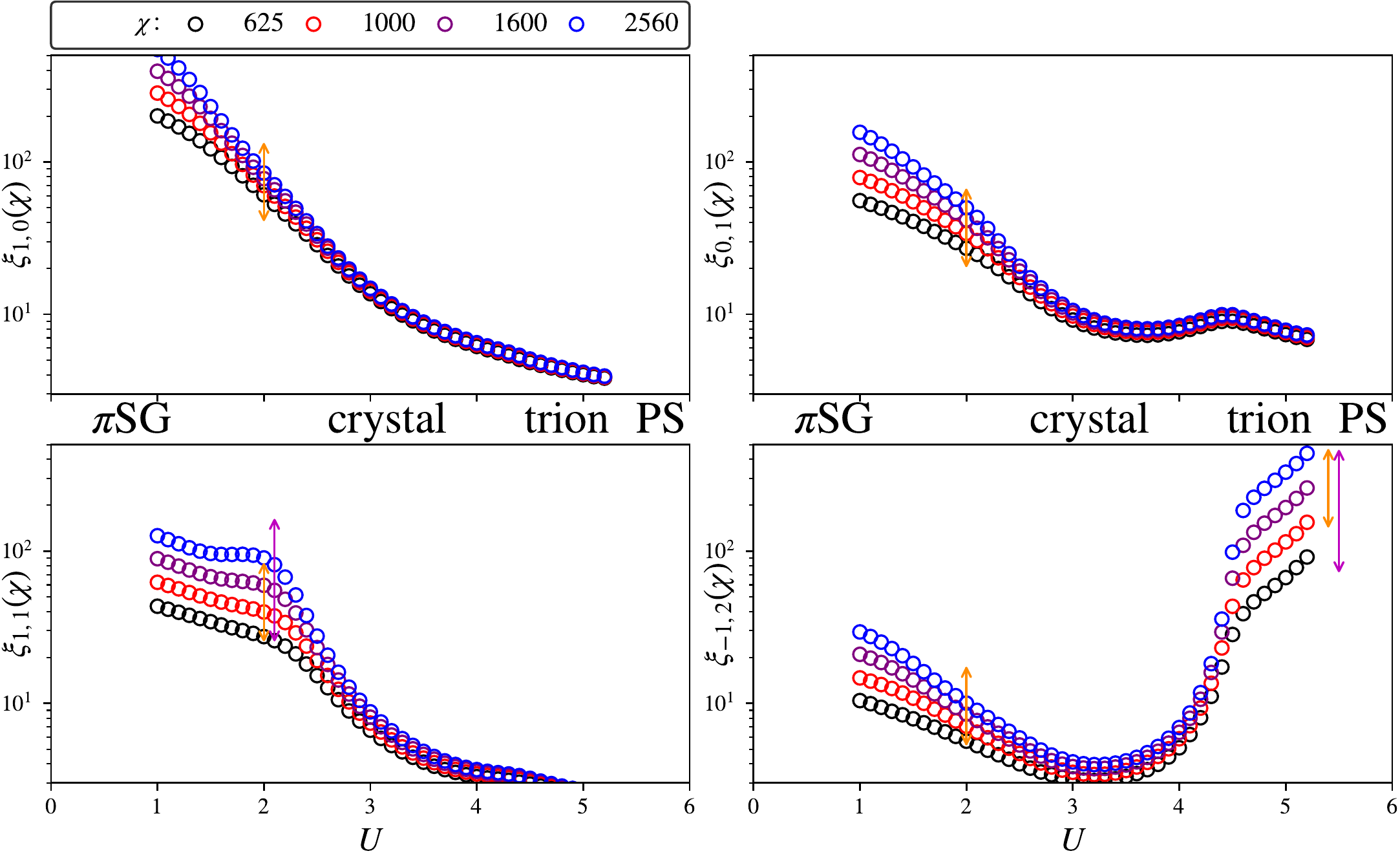}
\\
\caption{The phases of the model at 1/3 filling inferred from the finite-$\chi$ correlation lengths in $(1,0)$, $(0,1$), $(1,1)$, and $(-1, 2)$
sectors. The data is taken for different interaction $U$ along the cut $\alpha=\frac{t_a-t_b}{t_a+t_b}=0.9$. This cut shows the evidence of  $\pi \mathrm{SG}$, crystal and trion phase from left to right. (`PS' denotes phase separated.) For the four correlation lengths plotted, a $\pi \mathrm{SG}$ state is characterized by that only $\xi_{1,1}$ is divergent;  a trion state only has divergent $\xi_{-1,2}$ among the four sectors, and all lengths are finite for crystal. The philosophy of using a set of finite-$\chi$ correlation lengths to infer the exact values is the same as Fig.~\ref{fig:xi}. The solid orange and magenta lines represent the proposed increment from $\ln (\xi(\chi=625))$ (black circle) to $\ln(\xi(\chi=2560))$ (blue circle) in a gapless sector of $c=2$ and $c=1$ systems respectively, assuming no intervening from the gapped degree of freedom. }
\label{fig:onethirdxi}
\end{figure*}

For other commensurate fillings, the scaling dimension analysis shows that the locking of any umklapp term requires finite strength of interaction, indicating possibly richer phase diagrams. In the following paragraphs, we  revisit the phase diagram at 1/3 filling, studied previously in Ref.~\onlinecite{roscilde2012pairing}. In contradiction to Ref.~\onlinecite{roscilde2012pairing}, we argue for the existence of a $\pi \mathrm{SG}$ phase using both numerical evidence and theoretical arguments.

We use DMRG correlation lengths analysis, similar to Sec.~\ref{pilocking}, to construct the phase diagram at 1/3 filling, see Fig.~\ref{fig:phase diagram}(b). A sample of the DMRG correlation length data, taken for a cut at fixed $\alpha=0.9$ and varying $U$ is plotted in Fig.~\ref{fig:onethirdxi}. We observe that for intermediate interactions (centered on $U=3$), all four correlation lengths $\xi_{1,0}(\chi)$, $\xi_{0,1}(\chi)$, $\xi_{1,1}(\chi)$ and $\xi_{-1,2}(\chi)$ tend to converge, consistent with a crystal phase. For small $U$, we observe that $\xi_{1,1}(\chi)$ tends to diverge while the other correlation lengths tend to converge, which is consistent with the $\pi \mathrm{SG}$ phase. On the other hand, for large $U$, $\xi_{-1,2}(\chi)$ is the only correlation length that tends to diverge, indicating that the elementary bulk gapless excitation carries charge  $(-1,2)$, which is a trion composed of an a-hole and two b-particles.  The trion and crystal phases have been found previously in Ref.~\onlinecite{roscilde2012pairing}. Our correlation length data further confirms the trion phase by showing the elementary gapless excitation is trion.  The main difference between the conclusion of Ref.~\onlinecite{roscilde2012pairing} and our data (Fig.~\ref{fig:onethirdxi}) is that the ``missing" $\pi \mathrm{SG}$ order is indeed not missing. We will see from the theoretical analysis below that $\pi \mathrm{SG}$ is expected to neighbor the crystal phase with the trion phase on the latter's other side.

To interpret the data we obtained, it is sufficient to limit our attention to the most relevant umklapp term and the spin locking term. We first analyze the candidates and pick out the most relevant umklapp term. We observe that the resulting term is consistent with our finding of a trion phase.  Adjacent to the non-interacting limit,  the most relevant umklapp locking terms at 1/3 filling are: $g_{2,4}\cos(2\phi_a+4\phi_b)$ and $g_{4,2}\cos(4\phi_a+2\phi_b)$ where $g_{2,4}<0$ and $g_{4,2}<0$ for repulsive interaction.  With asymmetry of $a$ and $b$ due to mass imbalance, it is possible to have one of the two terms locked. To estimate which term is more relevant, we use bosonization to evaluate the scaling dimension of the two terms in the free theory. In Appendix~\ref{scaling dimension}, using the ``naive" parameters for the effective theory, we find that $\Delta_{(4,2)\text{pt}}$, the scaling dimension of  $g_{4,2}\cos(4\phi_a+2\phi_b)$, is smaller.  This term can be interpreted as binding two heavy holes with one light particle or equivalently two heavy particles with one light hole. This is because the dual field $-\theta_a+2\theta_b$, corresponds to $(-1,2)$ charge if appearing at the exponents of vertex operators, commutes with $4\phi_a+2\phi_b$ and thus is an independent field and  remains gapless after the latter gets locked. ($[\theta_{\sigma}(x,t),\phi_{\sigma^{\prime}}(y,t)]=i\pi \delta_{\sigma,\sigma^{\prime}} H(x-y)$, where $H$ is Heaviside step function.) Assuming only $g_{4,2}\cos(4\phi_a+2\phi_b)$ gets locked, the ground state is a liquid of trions with charge $(-1,2)$.  Recall from the last section, the locking of the spin boson results in the $\pi \mathrm{SG}$ phase. The locking of both the spin boson and $4\phi_a+2\phi_b$ gives a crystal phase, as the number of gapless modes is reduced to zero and long-range density order forms with three sites per unit cell. With the above picture in mind, we see that a direct transition from 2TLL to the crystal phase is unlikely, as there is no symmetry that induces the two locking terms to  lock simultaneously. 

In Appendix~\ref{spingapcrystal}, we provide additional data extracting ``spin gap" of open chains in presence of charge or trion gap; and  we discuss the remnants of filling anomaly.

\section{Conclusion and discussion}
We have studied the phase diagram of the repulsive one-dimensional Hubbard model with mass imbalance. We find $\pi \mathrm{SG}$ phase as the ground state of incommensurate fillings and study its quasi-long-range orders and string orders via bosonization and DMRG.  We point out $\pi \mathrm{SG}$ locking leads to filling anomaly~\cite{PhysRevB.99.245151, khalaf2019boundary} of open chains. For the equal-filled sector, the observed states do not have a spin gap and spontaneous breaking of inversion symmetry; the spin-gapped states are in the sector with one more heavy particle. We argue that this is made possible by the asymmetry between the two components in addition to the $\pi \mathrm{SG}$ locking.  The $\pi \mathrm{SG}$ phase is also shown to be a precursor phase of a type of crystal phases for commensurate fillings, which requires further particle-hole binding instabilities.  The ``spin gaps" of open chains of the crystals phases are also calculated and discussed.

Finally, we discuss some possible implications  of our results to the experiments using quantum simulators. Our calculations show that the amplitude of string order parameters could be much larger than the TSCz (``triplet" pairing) quasi-long-range order in certain parameter regions. Therefore, to experimentally detect the $\pi \mathrm{SG}$ phase we suggest measuring the string correlations~\cite{Hilker484}, instead of trying to detect TSCz order, and comparing the results with our predictions. Those observations can identify that mass imbalance drives the spins in squeezed space into N\'eel order. For the trion phase, the two components have different scaling dimensions for their density-wave orders; this difference may be detectable by Friedel oscillations.

\begin{acknowledgments}
We thank Erez Berg, Jennifer Cano, Juan Carrasquilla, Thierry Giamarchi, Ludwig Mathey, Frank Pollmann, Marcos Rigol, and  Ruben Verresen for discussion. This work used the Bridges system, which is supported by NSF award number ACI-1445606, at the Pittsburgh Supercomputing Center (PSC). This work was supported by the Charles E.\ Kaufman Foundation, NSF.\ DMR-1848336, NSF.\ PHY-1913034, and NSF.\ PIRE-1743717.
\end{acknowledgments}

\appendix

\section{Numerical methods}\label{DMRGnotes}
The first aspect is to determine the exact correlation lengths of the ground state in charge sectors $(q_a,q_b)$, which are denoted as $\xi_{q_a,q_b}$.

The ``finite-$\chi$ correlation lengths" are the correlation lengths of the best approximate ground state for an iMPS variational ansatz with bond dimension $\chi$. (The effectiveness of iDMRG can be checked against other iMPS optimization algorithms, e.g., Refs.~\onlinecite{VUMPS, iTEBD}.) We denote them  as $\xi_{q_a,q_b}(\chi)$. As the approximate state becomes exact with $\chi \rightarrow \infty$, the DMRG correlation lengths approach the exact values. We need to estimate if  $\xi_{q_a,q_b}(\chi)$ converges to a finite value or diverges based on finite-$\chi$ data. 
It is helpful to notice that if a correlation length is divergent,  the expected divergent rate is algebraic in $\chi$. Ideally, if the degrees of freedom of a state are decoupled CFTs, the rate to diverge is predictable in the limit $\chi \rightarrow \infty$~\cite{pollmann2009theory}:
\begin{subequations}
\begin{align}
\xi(\chi) &\propto \chi^{\kappa}\label{xiscaling},\\ \kappa &= \frac{6}{c(1+\sqrt{12/c})}\label{kappa},    
\end{align}
\end{subequations}
where $\xi$ denotes any divergent correlation length and $c$ is the sum of the central charge of each CFT.  Extended TLL phases are effectively described by decoupled free boson CFT(s), each with central charge 1. The TLL may also contain gapped degree(s) of freedom, 
as the consequence of locking some mode(s) of its parent theory. In this case, it is unclear that if  Eq.~\eqref{kappa} holds for $\chi \rightarrow \infty$. However, it is expected that $\kappa(c_{p}) < \kappa\leq\kappa(c)$, where $c_{p}$ is the total central charge of the parent theory~\footnote{It follows from that the entanglement spectrum of a gapped bosonic mode decays faster than a gapless bosonic mode}.

The method of using numerical data and Eq.~\eqref{xiscaling} to infer if $\xi_{q_a,q_b}$ is infinite is as follows. For a $\xi_{q_a,q_b}$, we estimate a value of $\kappa$ use a series of finite $\chi$ by assuming the form of Eq.~\eqref{xiscaling}. In the $\chi \rightarrow \infty$ limit, if $\xi_{q_a,q_b}$ is finite, the estimated $\kappa$ is by definition 0 , otherwise it follows Eq.~\eqref{kappa}.  As the limit $\chi \rightarrow \infty$  cannot be reached numerically,   we analyze data to see if the algebraic divergence relation is violated.  It is helpful to pick a geometric series of $\chi$ (e.g., [625, 1000, 1600, 2560]) to implement iDMRG and analyze the logarithm of the obtained correlation lengths. In this case, the  data of $\xi_{q_a,q_b}(\chi)$ is supposed to be equally spaced assuming a constant algebraic increase rate.  If $\xi_{q_a,q_b}$ is finite and is not much larger than $\xi_{q_a,q_b}(\chi=2560)$,  we expect to see decreasing spacing for increasing $\chi$, showing convergence.  For a TLL phase with some finite $\xi_{q_a,q_b}$, there must be other divergent $\xi_{q_c,q_d}$.  If the exact $\xi_{q_a,q_b}$ is not much larger than and not too close to $\xi_{q_a,q_b}(\chi=2560)$, the set of $\xi_{q_c,q_d}(\chi)$ can display crossover behavior, which is  increasing spacing for increasing $\chi$. The reason is that for small $\chi$ [$\xi_{q_a,q_b}(\chi)<<\xi_{q_a,q_b}$], the system behaves like its parent theory with a larger central charge, and thus a smaller $\kappa$ [Eq.~\eqref{kappa}] leading to smaller spacing. The  challenge to detect a small gap is that the implemented $\chi$ could be small in the sense that any finite $\xi_{q_a,q_b}$ is much larger than $\xi_{q_a,q_b}(\chi=2560)$. In this case, the above finite-$\chi$ analysis will not show a signal of a gap. In the main text and Appendix~\ref{BKT scaling collapse}, we use scaling dimension estimation and scaling collapse to make the detection less challenging.

The second aspect is using ``DMRG finite-$\chi$ density-wave order" to infer the exact density-wave quasi-long-range orders or long-range orders.

The TLL phases we study have density-wave quasi-long-range orders while the crystal phases have long-range density-wave orders. 
Here we discuss the finite-$\chi$ scaling used to distinguish quasi-long-range orders from long-range orders as well as extract wavevectors and exponents of quasi-long-range orders.   In our case, the observed phases of this model all have density quasi-long-range orders or long-range orders with a single base period $p$, compatible with our MPS ansatz with $p$ site per unit cell, where $p$ is the denominator of the irreducible filling fraction $q/p$.  With such compatibility, there is a finite-$\chi$ scaling for quasi-long-range orders represented by charge (0,0) operator $\rho$:

\begin{align}\label{quasiorderscaling}
\left\langle \rho(k) \right\rangle_{\chi} \propto \xi^{-\Delta_{\rho(k)}}_{0,0}(\chi),
\end{align}
where $\rho(k)$ is the Fourier transform of $\rho(x)$ over one unit cell: $\rho(k)=\sum_{x=1}^{p}e^{ikx}\rho(x)$, $\Delta_{\rho(k)}$ is the scaling dimension the density-wave quasi-long-range order of $\rho$ at wavevector $k$. This is to say, any finite-$\chi$ approximation of a (uniform) TLL ground state is not uniform in terms of $\rho$,  but get close to the uniform limit in a given manner with increasing $\chi$~\cite{emtll}~\footnote{An intuitive explanation is as follows: DMRG  density correlation of $\rho$ is faithful only within the length $\sim \xi_{0,0}(\chi)$; beyond this length scale, the DMRG result behaves like a mean-field theory, which has artificial long-range order compatible with the ansatz.}.  Equation~\eqref{quasiorderscaling} provides a finite-$\chi$ scaling method to extract density quasi-long-range order, which is practically more convenient and accurate than directly analyzing the DMRG correlation function. Besides quasi-long-range order, for an operator $\rho$ with true long-range density-wave order at $k$, $\rho(k)_{\chi}$ does not decay to zero.  Without any order at $k$, $\rho(k)_{\chi}$ exponentially decays to zero.  We remark that the scaling Eq.\eqref{quasiorderscaling} also applies if $\rho$ is a string operator, where the square of Eq.\eqref{quasiorderscaling} can be numerically evaluated.

\section{Inversion symmetry of \texorpdfstring{$\Phi_z(x)$ and $\Phi_n(x)$}{Phi[z] and Phi[n]} }\label{spatialparitystring}

Recall from the main text, $\Phi_n$ and $\Phi_z$ are defined as:

\begin{align}\label{recallJz}
\begin{split}
\Phi_{n}(x) &\equiv \left[\prod_{j<x}(-1)^{n(j)}\right](1-n(x))\\
&=\frac{1}{2}\left[\prod_{j<x} Q(j)\right](Q_a(x)+Q_b(x)),\\
\Phi_{z}(x) &\equiv \left[\prod_{j<x}(-1)^{n(j)}\right]\sigma_{z}(x)\\
&=\frac{1}{2}\left[\prod_{j<x} Q(j)\right](Q_a(x)-Q_b(x)).
\end{split}
\end{align}

We first discuss the inversion symmetry of $\Phi_n$ and $\Phi_z$.  Consider a quantum state $|I \rangle$ with inversion symmetry and its total fermion parity is $P$. Consider an inversion symmetric pair of of positions  $x$ and $x^{\prime}$, we claim that 
\begin{align}\label{stringparity}
\begin{split}
&\left\langle I| \Phi_{n}(x)| I \right\rangle=(-1)^{P}\left\langle I |\Phi_{n}(x^{\prime}) |I \right\rangle,\\  
&\left\langle I| \Phi_{z}(x)| I \right\rangle=-(-1)^{P}\left\langle I |\Phi_{z}(x^{\prime}) |I \right\rangle.
\end{split}
\end{align} 
The proof is as follows:

Multiply the total fermion parity operator $P=\Pi_{j}Q(j)$ to  $\Phi_n(x)$ and $\Phi_z(x)$:
\begin{align}\label{stringparityp1}
\begin{split}
&P\Phi_{n}(x)=\frac{1}{2}(Q_a(x)+Q_b(x)) \prod_{j>x} Q(j),\\
&P\Phi_{z}(x)=-\frac{1}{2}(Q_a(x)-Q_b(x)) \prod_{j>x} Q(j).
\end{split}
\end{align} 
We have used $Q(x)Q_{a}(x)=Q_{b}(x)$ and $Q(x)Q_{b}(x)=Q_{a}(x)$. Consider a state $|I \rangle$ that is symmetric (even or odd) under inversion which takes $x$ to $x^{\prime}$

\begin{align}\label{stringparityp2}
\begin{split}
&\left\langle I| \frac{1}{2}(Q_a(x)+Q_b(x)) \prod_{j>x} Q(j)| I \right\rangle = \left\langle I| \Phi_{n}(x^{\prime})| I \right\rangle,\\
&\left\langle I| \frac{1}{2}(Q_a(x)-Q_b(x)) \prod_{j>x} Q(j)| I \right\rangle = \left\langle I| \Phi_{z}(x^{\prime})| I \right\rangle.
\end{split}
\end{align} 
We also have:
\begin{align}\label{stringparityp3}
\begin{split}
&\left\langle I| P\Phi_{n}(x)| I \right\rangle=(-1)^{P}\left\langle I |\Phi_{n}(x)) |I \right\rangle,\\  
&\left\langle I| P\Phi_{z}(x)| I \right\rangle=(-1)^{P}\left\langle I |\Phi_{z}(x)) |I \right\rangle.
\end{split}
\end{align} 
Equation~\eqref{stringparity} follows from Eqs.~\eqref{stringparityp1}-~\eqref{stringparityp3}.

\section{Bosonization convention and boson fields as pseudoscalars }\label{bosonizationsymmeries}

We first review some aspects of the bosonization of two-component fermions. 

In our convention, the bosonization representation of fermionic operators are

\begin{align}\label{fermionbosonization}
c_\sigma(y)\sim \sum_{j} \kappa_{\sigma,\text{sgn}(j+0.5)} e^{i[\theta_{\sigma}+(2j+1)(\phi_{\sigma}+k_{\text{F},\sigma}x)]}, 
\end{align}
where $\sigma=a,b$ is the component index, $\kappa_{\sigma,\pm}$ is the Klein factor and we have neglected coefficient before each term.

In this paper, we focus on $k_{\text{F},a}=k_{\text{F},b}$ up to  a caveat that $(k_{\text{F},a}-k_{\text{F},b})L$ is finite even in the thermodynamic limit, where $L$ is the system size. Considering this caveat, we adopt the following basis transform:
\begin{align}\label{chargespinappendx}
&\text{charge boson}: \phi_c=\frac{1}{\sqrt{2}}(\phi_a+\phi_b),\\
&\text{spin boson}:\phi_s=\frac{1}{\sqrt{2}}[\phi_b-\phi_a+(k_{\text{F},b}-k_{\text{F},a})x].
\end{align}
This convention reduces to Eq.~\eqref{chargespin} in the limit $k_{\text{F},a}=k_{\text{F},b}$.  The reason to choose this convention is that as we will show,  $(k_{\text{F},a}-k_{\text{F},b})x$ is not the correct background field configuration to be subtracted in case of that there is a spin gap. [The background field configuration subtraction is implicit by the construction Eq.~\eqref{fermionbosonization}.]

In this convention, the bosonization representation form  of $\sigma_z$ is:
\begin{align}\label{bosonizationsigmazsimple}
\sigma_z(x)=\frac{\sqrt{2}}{\pi}\partial_x \phi_s +...
\end{align}
Equation.~\eqref{bosonizationsigmazsimple} is just a reproduction of Eq.~\eqref{bosonizationsigmaz} with the oscillatory and higher order terms neglected. We assume that in the system, $\sum_{j}\sigma_z(j)=M$.  In the bosonic representation, we have: $2\sqrt{2}\phi_s(x_{\mathrm{right}})-2\sqrt{2}\phi_s(x_{\mathrm{left}})=\int dx 2\sqrt{2}\partial_x\phi_s=2\pi M$. In discussing open chains, we can fix the convention $2\sqrt{2}\phi_s(x_{\mathrm{left}})=0$ for convenience~\cite{Berg1DTSCz}.

In this convention, the bosonization representation of inversion transform is $x \rightarrow x^{\prime}=2x_0-x$, $2\sqrt{2}\phi_s(x) \rightarrow 2 \pi M-2\sqrt{2}\phi_s(x^{\prime})$, $2\sqrt{2}\phi_c(x) \rightarrow -2\sqrt{2}\phi_c(x^{\prime})$, $\kappa_{\sigma,+} \leftrightarrow \kappa_{\sigma,-}$, where $x_0$ is the inversion center. This transform preserves the convention that $2\sqrt{2}\phi_s(x_{\mathrm{left}})=0$. The boson fields are pseudoscalars in the sense of above transform.  Bosons being pseudoscalars is indicated by Eq.~\eqref{bosonizationsigmazsimple} as $\sigma_z(x)$ in this system is a scalar.
The above analysis also indicates that  for an inversion symmetric state $| I \rangle$: 
\begin{align}\label{pscalar}
\left\langle I |2\sqrt{2}\phi_s(x)| I \right\rangle=\left\langle I |(2\pi M-2\sqrt{2}\phi_s(x^{\prime}))| I \right\rangle, 
\end{align}

\section{Bosonization of \texorpdfstring{$\Phi_z(x)$ and $\Phi_n(x)$}{Phi[z] and Phi[n]} }\label{bosonizationapp}

With the above discussion in mind, we discuss the bosonization representation of $\Phi_z(x)$ and $\Phi_n(x)$ from the consideration of symmetry and commutation relation.

We first consider the symmetries of the operators and the symmetries of the Hamiltonian. 
The U(1) symmetry of the two operators dictates that the bosonization representation should not contain $\theta$ fields, as the $U(1)$ transform is  $\theta_{\sigma} + \delta \theta_{\sigma}$, with $\delta \theta_{\sigma}$ an arbitrary c-number.

Then we analyze the parity of the operators under the transform $a \leftrightarrow b$,  which leaves $\sigma_z \to -\sigma_z$, $n \to n$. If the Hamiltonian has the corresponding symmetry, the transform in bosonization representation is $\phi_s \rightarrow -\phi_s$ [otherwise no simple representation exists, see the discussion for Eq.~\ref{bosonizationn} and \ref{bosonizationsigmaz} in the main text.].  In this case, we conclude $\Phi_z$ ($\Phi_n$) should be odd (even) under $\phi_s \rightarrow -\phi_s$.
For the case of absence of inter-component symmetry, we  show that the form of the leading terms of $\Phi_z$ and $\Phi_n$ should keep invariant by considering the inversion symmetry.

We have figured out in the Appendix~\ref{spatialparitystring} that the spatial parity of $\langle \Phi_n(x) \rangle$ and $\langle \Phi_z(x) \rangle$ on an inversion symmetric state depends on the total fermion parity $P$. We also have introduced $M$ as the total polarization in Appendix~\ref{bosonizationsymmeries}, when $M$ is even (odd), $P$ is even (odd) and  $\langle \Phi_n(x) \rangle$ , $\langle \Phi_z(x) \rangle$ are respectively even (odd) and odd (even). Further consider  Eq.~\eqref{pscalar}, we conclude that the bosonization representation $\Phi_z(\phi_s, \phi_c)$ should obtain a factor $-(-1)^M$ under the transform $2\sqrt{2}\phi_s \rightarrow 2\pi M-2\phi_s$ and  $\phi_c \rightarrow -\phi_c$. Similarly, $\Phi_n(\phi_s, \phi_c)$ should obtain a factor $(-1)^M$ under the transform $2\sqrt{2}\phi_s \rightarrow 2\pi M-2\phi_s$ and  $\phi_c \rightarrow -\phi_c$.

Finally, we consider the commutation relations of the string operators with the fermion  $c_a(y)$ and $c_b(y)$. We see that $\Phi_z(x,t)$ anti-commutes (commutes) with $c_{a}(y,t)$, $c_{b}(y,t)$ on its left (right). 
Recall from Eq.~\ref{fermionbosonization}, $c_\sigma(y)\sim \sum_{j} \kappa_{\sigma,\text{sgn}(j+0.5)} e^{(i[\theta_{\sigma}+(2j+1)(\phi_{\sigma}+k_{\text{F},\sigma}x)])}$. We recall that $[\theta_{\sigma}(x,t),\phi_{\sigma^{\prime}}(y,t)]=i\pi \delta_{\sigma,\sigma^{\prime}} H(x-y)$, where $H$ is Heaviside step function. As a result, to guarantee the commutation relation, each term in the bosonization representation  should contain a vertex with odd coefficients before $\phi_a$ and $\phi_b$, i.e., $~\sim e^{i[(2j+1)(\phi_a+k_{\text{F},a}x)+(2l+1)(\phi_b+k_{\text{F},b}x)]}$, where $j, l\in \mathbb{Z}$.

With the consideration above and count the leading (lowest harmonic) term with no short wavelength oscillation, we find $\sin(\sqrt{2}\phi_{s})$, $\cos(\sqrt{2}\phi_{s})$ respectively for $\Phi_z$ and $\Phi_n$. Notice that $\Phi_z(x)$ becomes $\sigma_z$ at the left edge and $-(-1)^{P} \sigma_z$ at the right edge. In fact, the above consideration by itself rules out the possibility that $\cos(\sqrt{2}\phi_{s})$ as a term for $\Phi_z$. Similarly $\Phi_n(x)$ becomes $1-2n$ at the left edge and $(-1)^{P}(1-2n)$ at the right. These could be encoded by some of the subleading terms. We speculate that:

\begin{align}\label{bosonizationJzA}
\begin{split}
\Phi_{z}(x) \sim &\sin(\sqrt{2}\phi_{s})+\partial_x\phi_{s}(\cos(\sqrt{2}\phi_c+2k_{\text{F}}x)+\cos(\sqrt{2}\phi_{s}))\\
&+\sin(\sqrt{2}\phi_{s})\cos(2\sqrt{2}\phi_c+4k_{\text{F}}x)+...
\end{split}
\end{align}

\begin{align}\label{bosonizationJnA}
\begin{split}
\Phi_{n}(x) \sim &\cos(\sqrt{2}\phi_{s})+\cos(\sqrt{2}\phi_c+2k_{\text{F}}x)+\partial_x \phi_c\cos(\sqrt{2}\phi_{s})\\
&+\cos(\sqrt{2}\phi_{s})\cos(2\sqrt{2}\phi_c+4k_{\text{F}}x)+...
\end{split}
\end{align}

\section{BKT scaling collapse}\label{BKT scaling collapse}

\begin{figure}
\includegraphics[width=\columnwidth]{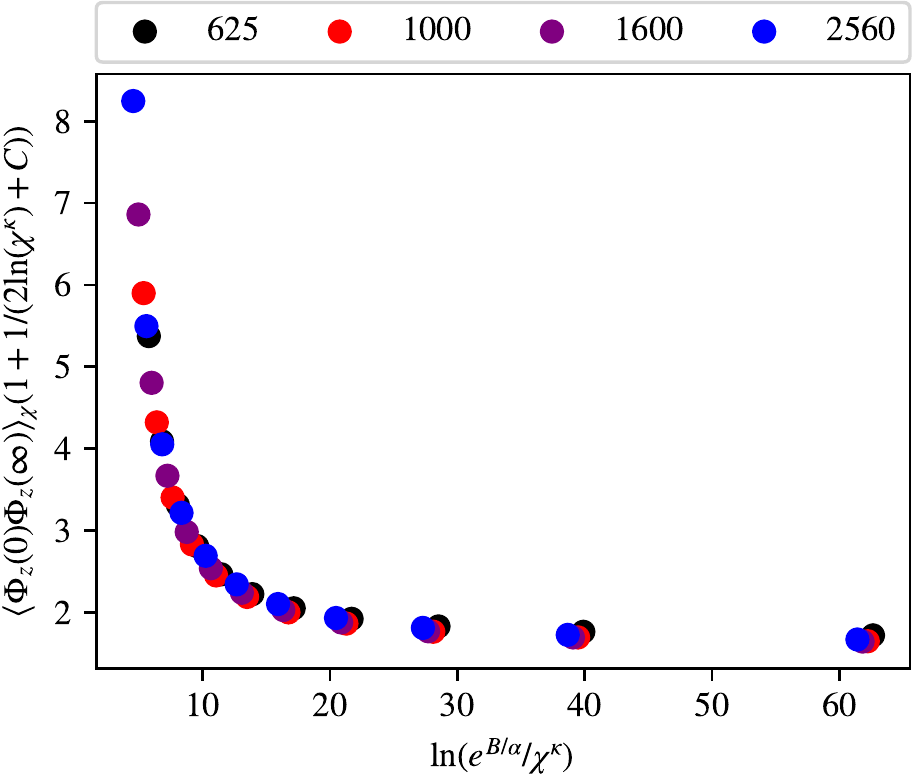}
\caption{A finite-$\chi$ BKT scaling collapse of order parameters. The BKT transition between 2TLL and $\pi \mathrm{SG}$ is assumed to happen at $\alpha=0$ and the order parameters scaling of infinite system is assumed to be Eq.~\eqref{gapscaling}. $B$ and $C$ are the fitting parameters to minimize the square residual. The data is taken with parameters: $U=3$, $0<\alpha=\frac{t_a-t_b}{t_a+t_b} \leq 0.5$.}
\label{fig:BKTscalingcollapse}
\end{figure}

 We employ a finite-$\chi$ scaling ansatz of an infinite system by substituting the system size of the  finite-size scaling~\cite{PhysRevA.87.043606, PhysRevB.91.165136} with the effective length scale $\chi^{\kappa}$.
As a proxy for the bulk gap $E_{\text{sg}}$, we can choose either the inverse single particle correlation length(s) or the string order ($\left\langle \Phi_{z}(0) \Phi_{z}(\infty)\right\rangle_{\chi}$). Here we present the result of string order collapse based on the assumption Eq.~\eqref{gapscaling}.  The critical point is assumed to be $\alpha=0$. We minimize the mean square residual to ``collapse" the scaled curves. The collapse result is shown in Fig.~\ref{fig:BKTscalingcollapse}. Despite the result shows no inconsistency with the assumption Eq.~\eqref{gapscaling}, more work is needed to draw a conclusion with a determination of $B$ in equation  Eq.~\eqref{gapscaling}. In particular, it may be helpful to consider  analyses similar to that of Ref.~\onlinecite{Hsieh_2013}.

\section{DMRG data for the crystal phase at half filling}\label{crystalhalffilling}
\begin{figure}
\includegraphics[width=\columnwidth]{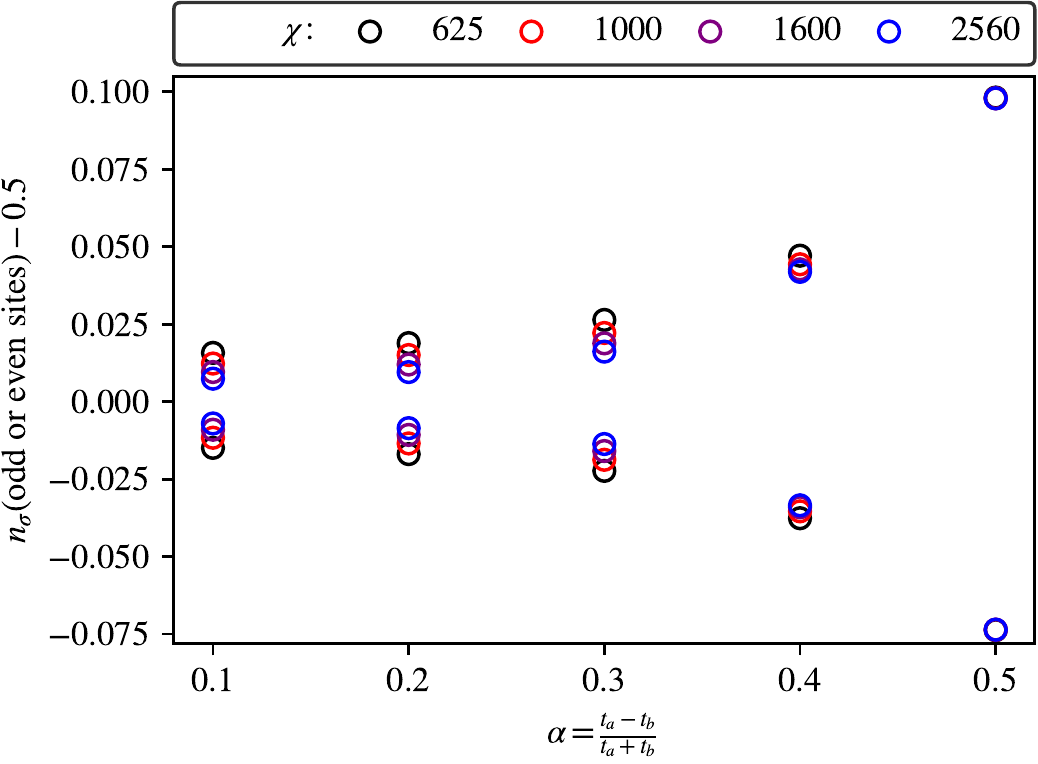}
\caption{Density-wave order of the ground states of the model at half filling.
The crystal phase has two sub-lattices (``even" and ``odd" ) with different densities $n_{\sigma}$, $\sigma=a,b$. We plot $n_{\sigma}$ (extracted by 0.5) of one of the sub-lattices. The $n_a-0.5$ is positive while  $n_b-0.5$ is negative; we see that the density waves of component $a$ and $b$ differ by a phase $\pi$. We also see that the amplitudes of density waves are different.  The finite-$\chi$ approximation works less accurate for small $\alpha$, i.e., those points are less overlapped with each other.
The interaction parameter is $U=3$.    }
\label{fig:onehalfDW}
\end{figure}

In Fig.~\ref{fig:onehalfDW}, we present DMRG data of  density-wave order for the model at half filling. The deviation of density from the average density is the order parameter of crystal.   For long-range order, unlike  quasi-long-range order (e.g., Fig.~\ref{fig:densityorder}), the order remains finite in the $\chi \rightarrow \infty$ limit. From Fig.~\ref{fig:onehalfDW} alone, it is not clear if the order is non-vanishing for small $\alpha$.

\section{Estimating scaling dimensions of vertex operators of the free theory}\label{scaling dimension}

We first write the effective theory with ``naive" parameters:
\begin{align}\label{Hfree}
\mathcal{H}_{\text{free}}=\sum_{\sigma}\frac{v_{\sigma}}{2}\left[\pi(\Pi_{\sigma})^2+(\partial_{x}\phi_{\sigma})^2/\pi \right]+\frac{U}{\pi^2}\partial_{x}\phi_{a}\partial_{x}\phi_{b},
\end{align}
where $v_{\sigma}=2t_{\sigma}\sin(k_{\text{F}})$. In our convention, $v_a>v_b$. As introduced in the main text, the most relevant locking terms are $g_s\cos(2\phi_a-2\phi_b)$; umklapp terms $g_{2,4}\cos(2\phi_a+4\phi_b)$, $g_{4,2}\cos(4\phi_a+2\phi_b)$ for 1/3 filling only.
\begin{figure}[htb]
\includegraphics[width=\columnwidth]{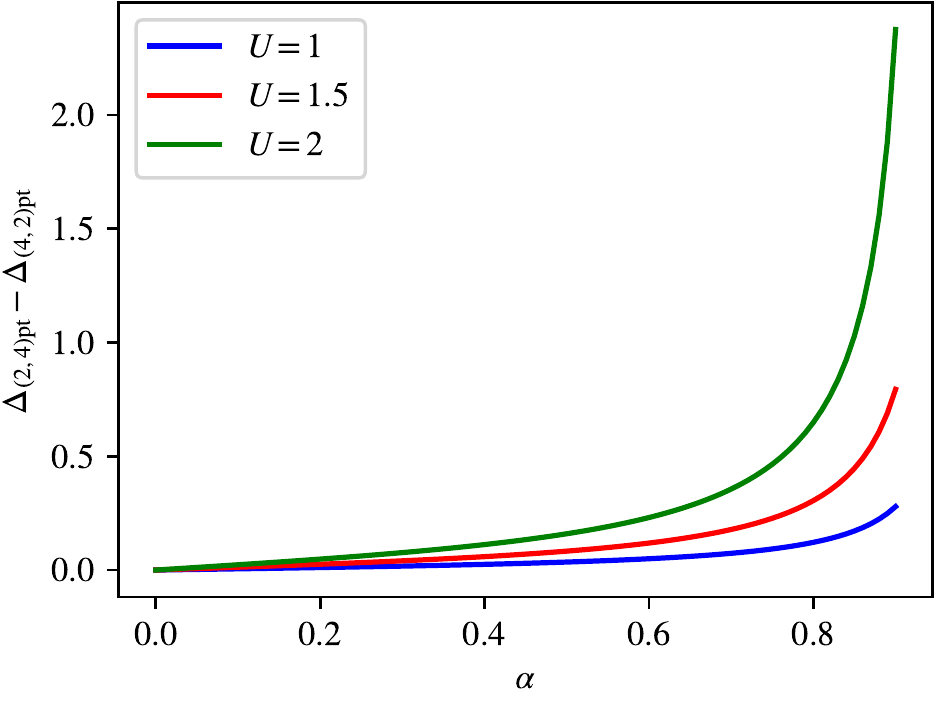}
\caption{The comparison of ``engineer" scaling dimensions of two vertex terms. For all the cuts plotted, $\Delta_{(4,2)\text{pt}}$ is smaller, and  the corresponding vertex term is likely to be more relevant.}
\label{fig:scalingdimensionfreetheory}
\end{figure}
Our goal is to estimate and compare the umklapp terms' scaling dimensions at the free theory:  $\Delta_{(2,4)\text{pt}}$ and $\Delta_{(4,2)\text{pt}}$.
To do this, we find the basis of two boson modes ($\phi_1$, $\phi_2$) such that they are decoupled in Eq.~\eqref{Hfree} with each other and their Luttinger parameters are 1.
\begin{align}
\left(                 
  \begin{array}{ccc}   
    \phi_{a}\\ 
    \phi_{b} \\ 
  \end{array}
\right)      
=
\left(                 
  \begin{array}{ccc}   
    \alpha_{1}&\alpha_{2} \\ 
    \beta_{1}&\beta_{2} \\ 
  \end{array}
\right) 
\left(                 
  \begin{array}{ccc}   
   \phi_{1} \\ 
   \phi_{2}\\ 
  \end{array}
\right),       
\end{align}

\begin{align}
\alpha_{1}^{4}=\frac{v_{a}+v_{b}d^2}{(v_{a}+v_{b}c^2+2Uc/\pi)(1-cd)^2},
\end{align}
\begin{align}
\beta_{2}^{4}=\frac{v_{a}c^2+v_{b}}{(v_{a}d^2+v_{b}+2Ud/\pi)(1-cd)^2},
\end{align}

\begin{align}
c \equiv \frac{\beta_{1}}{\alpha_{1}}=\frac{v_{a}^2-v_{b}^2-\sqrt{(v_{a}^2-v_{b}^2)^2+4U^2v_{a}v_{b}/\pi^2}}{-2Uv_{a}/\pi},
\end{align}
\begin{align}
d \equiv \frac{\alpha_{2}}{\beta_{2}}=\frac{v_{a}^2-v_{b}^2-\sqrt{(v_{a}^2-v_{b}^2)^2+4U^2v_{a}v_{b}/\pi^2}}{2Uv_{b}/\pi}.
\end{align}
With the new basis, we can evaluate 
\begin{align}
\Delta_{(2m,2n)\text{pt}}=\left[(m\alpha_1+n\beta_1)^2+(m\alpha_2+n\beta_2)^2\right]/4,
\end{align}
where the subscript $(2m,2n)$ labels $\cos(2m\phi_a+2n\phi_b)$. 
We plot $\Delta_{(2,4)\text{pt}}-\Delta_{(4,2)\text{pt}}$ for cuts of different $U$ in Fig.~\ref{fig:scalingdimensionfreetheory}. As the difference is always positive,  we infer that $\Delta_{(4,2)\text{pt}}$ is more likely to get locked. 
We note that Eq.~\eqref{Hfree} is only accurate in the weak-coupling limit under a certain regularization scheme, and a generic effective theory involves more terms. As it is difficult to know the parameters of the generic free theory as well as the coefficients of the vertex operators, we limit our analysis to the 0-loop estimation on the ``naive" free theory.

\section{The ``spin gap" of crystal phases}\label{spingapcrystal}

\begin{figure}[htbp]
(a) \raisebox{-0.9\height}{\includegraphics[width=8cm]{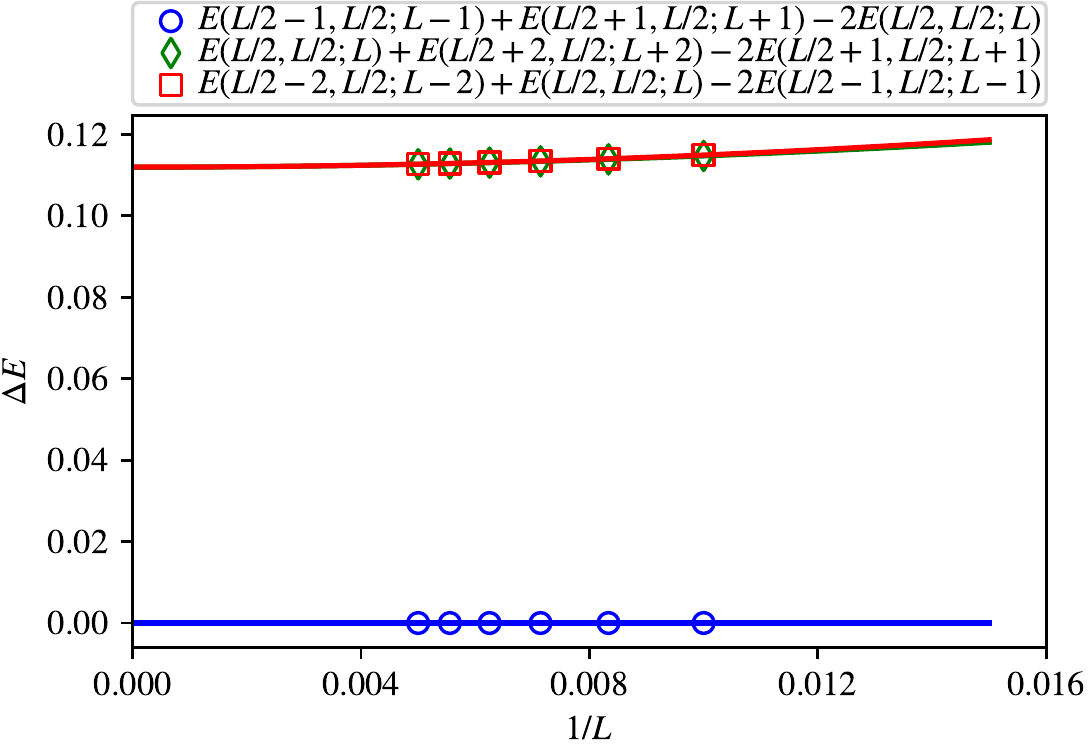}}
\\
(b) \raisebox{-0.9\height}{\includegraphics[width=8cm]{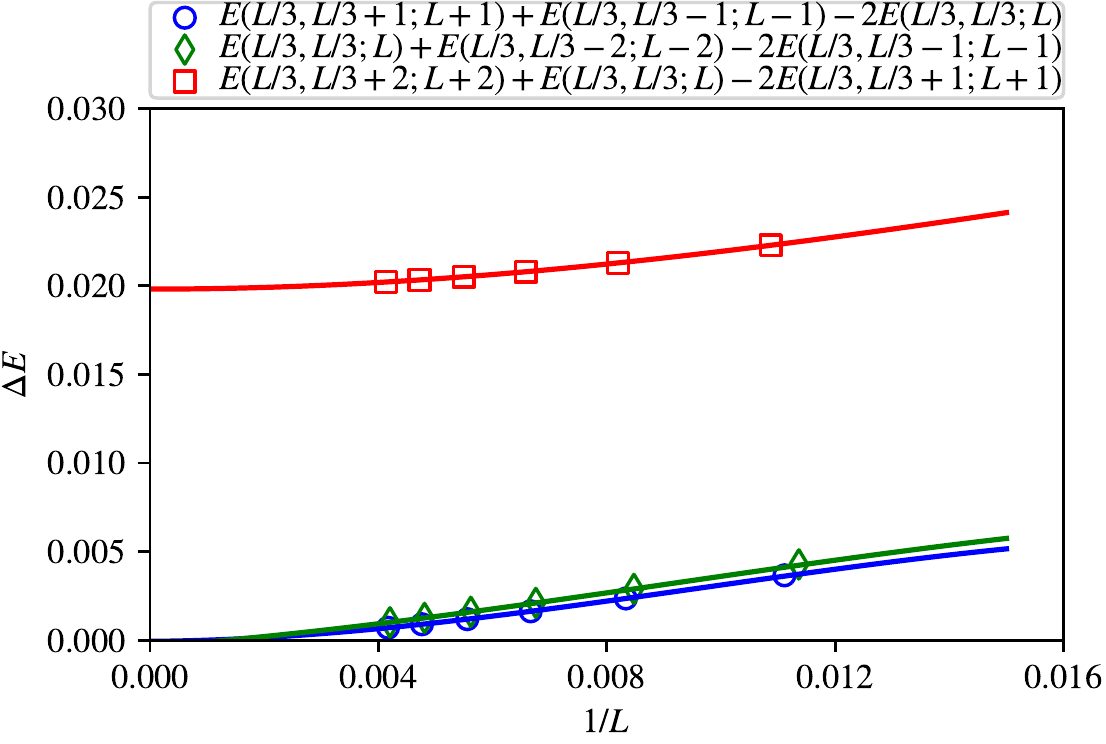}}
\caption{Extrapolating ``spin gap" of the  (a) half filling  and (b) one-third filling crystal phases.  For the labels, the definition follows the Fig.~\ref{fig:spingap} with caution that each $\Delta E$ is calculated by ground state energies of systems differ by lattice site numbers. For one-third filling, only the sector with one more heavy particle is ``spin-gapped", similar to observed incommensurate $\pi \mathrm{SG}$ states. The half filling shows different behavior.}
\label{fig:crystalspingap}
\end{figure}

First, we give a definition of the ``spin gap". The spin gap of $\pi \mathrm{SG}$ phase can be naturally defined from the summation of the energy cost of inserting and extracting one particle of a fixed type.  This definition cannot directly apply to the crystal phases because inserting and extracting one particle both make the commensurate criteria no longer hold. The criteria are: a. $N_a+N_b=L$ (1/2 filling); b. $2N_a+N_b=L$ (1/3 filling).  Therefore, not only the gap of spin boson but also the Mott gap or trion gap contribute to the gap value for the original definition.  To separate the spin gap with the Mott or trion gap, we keep the commensurate criteria by adjusting the system size $L$ while adjusting particle numbers. We increase (decrease) $L$ by the minimal possible number while adding (extracting) a particle.  

As the spin boson locks at the same value in these crystal phases as the $\pi \mathrm{SG}$ phase, there should be some remnants of  ``filling anomaly".
With this definition of the spin gap, our numerical results show that the one-half and one-third filling crystals show two distinct behaviors (Fig.~\ref{fig:crystalspingap}). The one-third filling crystal states only have a spin gap for systems with one more heavy particle.  For half filling, however,  systems with one more light particle are also spin-gapped. This behavior  is instead similar to that observed for $\pi \mathrm{SG}$ phase with additional spinful time-reversal symmetry~\cite{PhysRevB.81.064439}, while the definition of the  spin gap now involves varying the number of lattice sites.

As a remnant of  ``filling anomaly", the above energy landscape is related to inversion symmetry breaking.  We also observe from the density profiles that for the sector with each component being precisely half-filled, the ground states spontaneously break inversion symmetry~\footnote{Despite that the true ground state always preserves symmetry for any finite system, in this case,  there are symmetry broken states with energy exponentially close to ground state energy with increasing system size;  as long as the difference is smaller than numerical accuracy, DMRG calculation can stably find symmetry broken states. }; while  no spontaneous inversion symmetry breaking is observed for one-third filling and equally filled  ground states. This observation indicates that the spin configurations Figs.~\ref{fig:gsspinbosonconfig}(c) and (d)  are stable for half filling and is consistent that there is a bulk ``spin gap" for the precisely half-filled states.

\bibliography{main}
\end{document}